\documentclass[superscriptaddress,twocolumn,amsmath,amssymb]{revtex4-1}

\usepackage{preamble} 

\begin{document}

\title{Tunable RF-photonic filtering with high out-of-band rejection in silicon}

\author{Shai Gertler}
\email[]{shai.gertler@yale.edu}
\affiliation{Department of Applied Physics, Yale University, New Haven, CT 06520, USA}

\author{Eric A. Kittlaus}
\affiliation{Department of Applied Physics, Yale University, New Haven, CT 06520, USA}
\affiliation{Jet Propulsion Laboratory, California Institute of Technology, Pasadena, CA 91109, USA}

\author{Nils T. Otterstrom}
\affiliation{Department of Applied Physics, Yale University, New Haven, CT 06520, USA}

\author{Peter T. Rakich}
\email[]{peter.rakich@yale.edu}
\affiliation{Department of Applied Physics, Yale University, New Haven, CT 06520, USA}

\begin{abstract}
The ever-increasing demand for high speed and large bandwidth has made photonic systems a leading candidate for the next generation of telecommunication and radar technologies. The photonic platform enables high performance while maintaining a small footprint and provides a natural interface with fiber optics for signal transmission. However, producing sharp, narrow-band filters that are competitive with RF components has remained challenging. In this paper, we demonstrate all-silicon RF-photonic multi-pole filters with $\sim100\times$ higher spectral resolution than previously possible in silicon photonics. This enhanced performance is achieved utilizing engineered Brillouin interactions to access long-lived phonons, greatly extending the available coherence times in silicon. This Brillouin-based optomechanical system enables ultra-narrow ($3.5$ MHz) multi-pole response that can be tuned over a wide ($\sim10$ GHz) spectral band. We accomplish this in an all-silicon optomechanical waveguide system, using CMOS compatible fabrication techniques. In addition to bringing greatly enhanced performance to silicon photonics, we demonstrate reliability and robustness, 
necessary to transition silicon-based optomechanical technologies from the scientific bench-top to high-impact field-deployable technologies.
\end{abstract}
\maketitle

\section{\label{sec:intro}Introduction}
The seemingly endless appetite for high bandwidth, rapid reconfigurability, and high spectral resolution in modern communications is an impetus for new signal processing technologies that expand the capabilities offered by conventional RF circuits. One way to meet these challenges is to harness the complementary benefits offered by optical- and acoustic-wave signal processing technologies. Acoustic wave signal processing has long been a crucial part of modern RF systems \cite{aigner2008saw,ruppel2017acoustic}. Signal operations requiring narrow-band filtering and long delays invariably rely on electro-mechanical transduction to access slow-moving and long-lived acoustic waves, which are necessary to realize such operations within a small footprint \cite{aigner2018baw,balysheva2019saw}.

By comparison, electro-optic conversion of RF signals to the optical domain enables ultrawideband signal processing capabilities \cite{capmany2006tutorial,yao2009microwave,marpaung2019integrated}. In particular, rapidly evolving technologies based on silicon photonics offer reconfigurable wideband signal processing capabilities that leverage CMOS manufacturing techniques \cite{reed2004silicon_photonics_book,Jalali2006_Silicon_Photonics_review,thomson2016roadmap,komljenovic2016heterogeneous,shen2017deep,harris2017quantum,zhang2020photonic}. 
However, narrow-band ($\sim$MHz) filtering and long ($>100$ ns) signal delays --- of the type routinely performed in the acoustic domain --- are not yet possible in silicon photonics. 
Comparable performance in the optical domain demands ultralow-loss ($\sim0.1$ dB/m) waveguides and ultrahigh-$Q$ ($>10^8$) optical resonators \cite{little2004very,dong2010ghz,tien2011ultra,biberman2012ultralow,qiu2018continuously,onural2020ultra}, which are challenging to realize in silicon-photonic circuits.
Alternatively, if we could harness the performance benefits of acoustic-wave signal processing, while keeping signals in the optical domain, it could be possible to realize fully integrated silicon-photonic systems that may someday replace conventional microwave technologies.

Brillouin scattering, namely the coupling of light and acoustic phonons in the RF-frequency range, enables access to the narrow-band acoustic-wave properties within the large-bandwidth optical domain \cite{garmire2017perspectives,klee2018applications,eggleton2019brillouin}. Brillouin-based filters, sensors, and oscillators were first demonstrated using discrete-component optical fiber technologies \cite{zhu2007stored,zadok2007gigahertz,zadok2011stimulated,zadok2012random}, and more recently have been developed on chip through the development of Brillouin-active waveguides \cite{otterstrom2018silicon,zarifi2019chip,xie2019system,eggleton2019brillouin}. The long-lived acoustic waves that mediate Brillouin interactions yield narrow spectral features, similar to the role played by acoustic waves in RF filters \cite{aigner2008saw,ruppel2017acoustic}, with resonant frequencies in the microwave range, and are further tunable through optical wavelength and device geometry \cite{zadok2007gigahertz,sancho2012tunable,casas2015tunable}. These features make Brillouin-scattering-based devices a promising candidate for RF-photonic applications such as filters \cite{xie2019system}, delay lines \cite{zadok2011stimulated, merklein2018brillouin} and oscillators \cite{li2013microwave,shi2018generation}. Further, the recent demonstrations of Brillouin scattering in silicon \cite{van2015interaction,kittlaus2016large,kittlaus2017_Intermodal} could facilitate low-cost, high-volume production using CMOS-fabrication techniques, and enable the integration of photonic and electronic components on the same platform \cite{soref2006past,hochberg2010towards,atabaki2018integrating}.

A promising strategy to utilize Brillouin scattering for RF-filtering operations is a photonic-phononic emitter- receiver (PPER), where information is transduced be- tween spatially separated optical waveguides using sound waves, resulting in a narrow-band frequency response \cite{shin2015control,kittlaus2018rf,gertler2019microwave}. This is accomplished by encoding information on an optical carrier through intensity modulation, which drives spatially-extended coherent acoustic waves through a forward Brillouin interaction. The driven elastic deformation induces a narrow-band phase modulation onto light propagating in a separate optical waveguide, which after demodulation, results in a narrow RF pass-band filter response corresponding to the spectral properties of the acoustic modes. This concept has been demonstrated in integrated waveguides \cite{shin2015control,kittlaus2018rf} and optical fibers \cite{diamandi2017opto}. In contrast to the Lorentzian line shapes that are typical of Brillouin interactions, it is possible to couple multiple acoustic modes in a PPER scheme to produce a multi-pole filter response \cite{shin2015control,gertler2019shaping}. Recent theoretical studies indicate that RF links based on PPER devices can meet the performance requirements needed for practical applications \cite{ gertler2019microwave}. Experimental studies have explored the performance of static single-pole PPER filters \cite{kittlaus2018rf}, and devices fabricated using specialized MEMS fabrication techniques have shown excellent multi-pole frequency response \cite{shin2015control}. To have technological impact in real-world applications, PPER-based filters require a combination of frequency-tunable operation, excellent link performance, and robust foundry compatible fabrication \cite{marpaung2019integrated}.

In this work, we report a tunable narrow-band RF-photonic filter based on a multi-pole PPER device, as a basis for versatile new RF-photonic systems using standard silicon-on-insulator (SOI) fabrication methods. The PPER devices consist of suspended rib optical waveguides supporting an optical mode as well as a long-lived acoustic mode, resulting in strong forward Brillouin coupling of light and sound \cite{shin2013tailorable,kittlaus2016large}. We show how a phononic crystal design with an acoustic stop-band enables coupling of acoustic modes in a controllable fashion. The resulting acoustic multi-mode interference yields a second- order filter response, with 3.5 MHz full-width at half- maximum (FWHM) at a center frequency of 3.87 GHz, and 70 dB out-of-band suppression. Using this device, we demonstrate an RF-photonic link, yielding a link-gain of $G = -17.3$ dB. We further show that the addition of an RF amplifier at the link input achieves larger-than-unity gain ($G = 0.6$ dB) and improves the link noise-figure. Additionally, by introducing a tunable local oscillator, we demonstrate tunability of the filter pass-band over multiple GHz, while maintaining the highly selective, narrow-bandwidth filter shape. The RF-photonic link characteristics show good agreement with a first-principles model of link performance, serving as a foundation from which to build on this technology.
\begin{figure*}[htb!]
    \centering
    \includegraphics[scale=0.77]{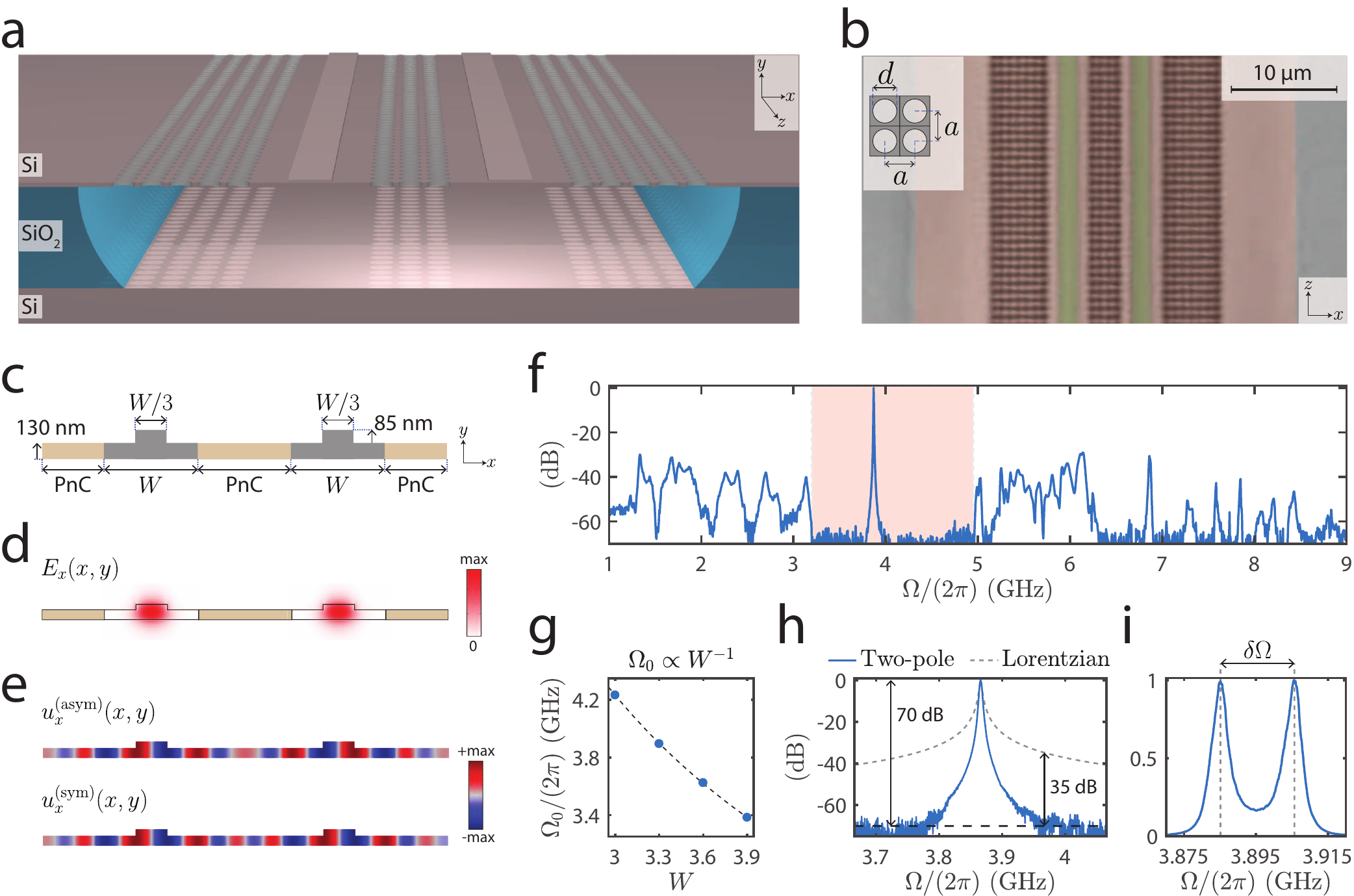}
    \caption{
    \textbf{(a)} Artistic representation of a two-pole PPER device, showing the two suspended rib waveguides surrounded by a phononic crystal (PnC) cubic lattice.
    \textbf{(b)} Micrograph image with enhanced colors of a top-down view of a two-pole PPER filter. 
    Inset, the cubic lattice is parameterized in terms of its pitch ($a$) and hole diameter ($d$).
    \textbf{(c)} Schematic diagram of the device cross-section denoting the ridge waveguide sections and the phononic-crystal (PnC) regions.
    \textbf{(d)} FEM simulation of the x-component of the optical modes $E_x$ in both rib waveguides.
    \textbf{(e)} FEM simulation of the displacement x-component of the  of the two acoustic super-modes of the suspended structure, $u^{(asym)}_x$, and $u^{(sym)}_x$ denoting the asymmetric and symmetric modes, respectively.
    \textbf{(f)} Measured frequency response of the device, revealing an acoustic stop-band with a defect mode around 3.9 GHz. The defect region width is $W=3.3\ \mu$m, and the phononic crystal parameters are $a=650$ nm and $d=500$ nm, with $N=5$ lines of holes between the two waveguides.
    \textbf{(g)} The measured acoustic resonant frequency is inversely proportional to the defect widths $\Omega_0\propto W^{-1}$.
    \textbf{(h)} Magnified view of the peak from panel (f) reveals a sharp frequency roll-off. A Lorentzian line shape with the same FWHM is shown for reference.
    \textbf{(i)} Measured response of a device with a strong coupling between the two acoustic modes, showing a frequency splitting corresponding to the two acoustic super-modes supported by the structure. The defect region width is $W=3.3\ \mu$m, and the phononic crystal parameters are $a=600$ nm and $d=462$ nm, with $N=3$ lines of holes between the two waveguides.
    }
    \label{fig:device}
\end{figure*}

\section{\label{sec:results}Results}
\subsection{\label{subsec:device}Device design}
We demonstrate the multi-pole filter response using the opto-mechanical device structure shown in Figs. \ref{fig:device}(a-c). The device is fabricated in a single-crystal silicon layer of a silicon-on-insulator (SOI) chip, as detailed in the Supplement, Section \ref{sec:SI-fab}. Two identical rib waveguides provide optical guiding for optical TE-like fields, shown in Fig. \ref{fig:device}(d). The waveguides are suspended by removing the underlying oxide, enabling long-lived acoustic modes to be guided in the silicon layer \cite{kittlaus2016large}. A cubic lattice of air holes defines a phononic crystal with two line-defects, such that acoustic modes are guided in the same spatial region as the optical fields. The rib design enables low-loss optical propagation, while the air under-cladding enables tight acoustic confinement, resulting in strong forward Brillouin coupling over the entire device length \cite{shin2013tailorable,kittlaus2016large}. The acoustic response resulting from the phononic crystal and defect regions is measured by modulating the intensity of an optical tone in one of the waveguides and detecting the Brillouin-induced phase modulation in the second waveguide \cite{ kittlaus2018rf,gertler2019shaping} (for more details see Section \ref{subsec:filter}). Fig \ref{fig:device}(f) presents the measured acoustic response, showing a stop-band for Brillouin-active phonon modes over a frequency range of 1.76 GHz, and a sharp peak at 3.9 GHz corresponding to the line-defects in the structure.

\begin{figure}[hb!]
    \centering
    \includegraphics[scale=0.77]{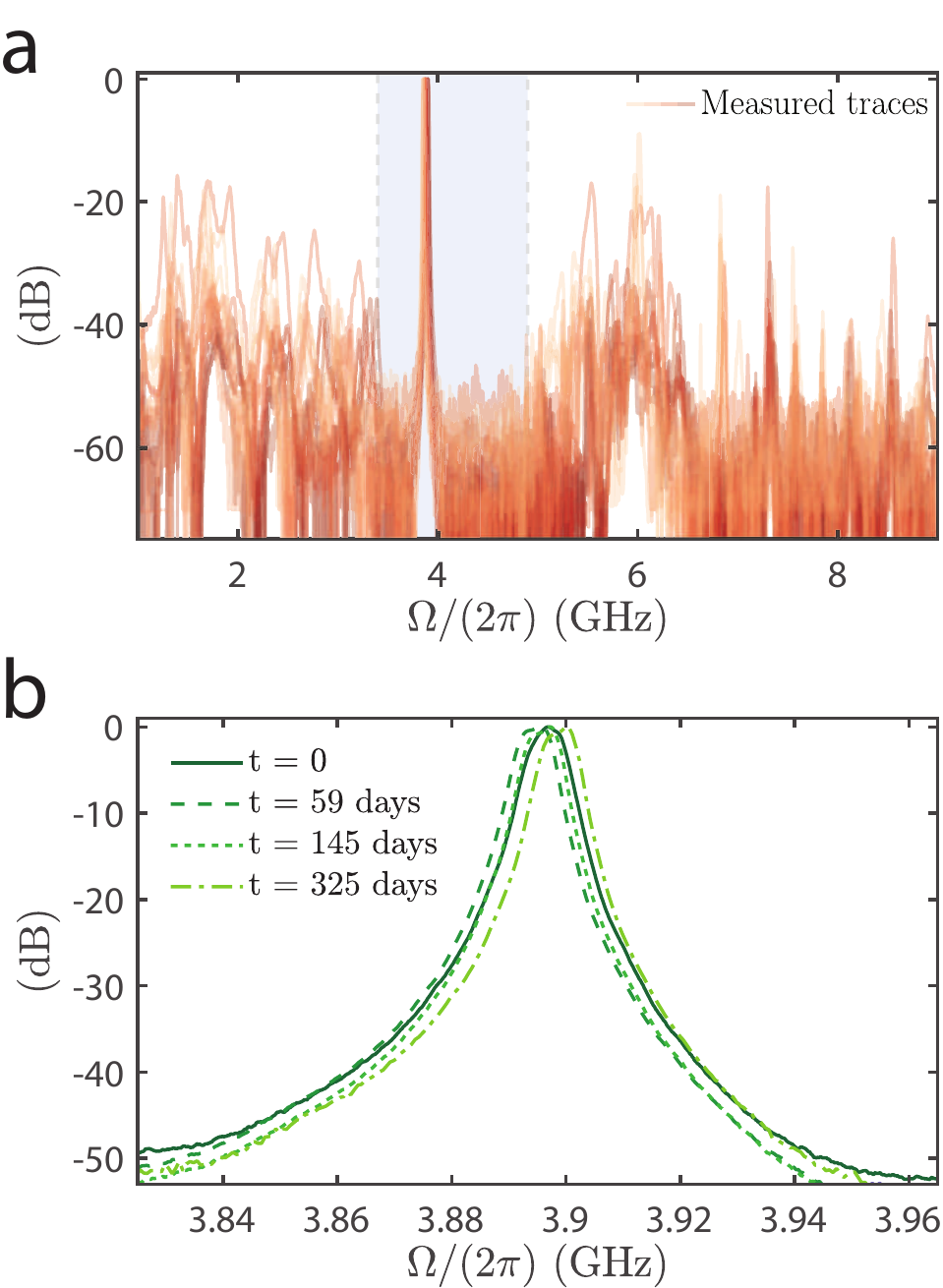}
    \caption{
    \textbf{(a)} Measurements of sixteen devices fabricated on a single chip, all showing a stop-band width on the order of 1.5 GHz and a defect mode close to 4 GHz.
    \textbf{(b)} Repeated measurements of a single device over a period of eleven months show no degradation of performance nor alteration of the two-pole line shape. The center frequency shows a fluctuation of 0.1\%, which can be the result of temperature variation between measurements.
    }
    \label{fig:fab_1}
\end{figure}

The use of two coupled acoustic modes enables the design of a second-order frequency response desirable for high out-of-band suppression, which is unattainable with a single acoustic resonance \cite{shin2015control,gertler2019shaping}. The phonon defect-modes are acoustically coupled through the region between the two waveguides, yielding symmetric and antisymmetric acoustic super-modes, as seen in Fig. \ref{fig:device}(e). The coherent interference of the two super-modes yields a sharp frequency roll-off, and high out-of-band suppression when comparing to a typical acoustic resonance which follows a Lorentzain line shape \cite{gertler2019shaping}, shown in Fig. \ref{fig:device}(h). The resonant frequencies of the two acoustic super-modes are separated by $2\mu$, where $\mu$ denotes the acoustic coupling rate. This results in a spectral line shape with two peaks, separated by a frequency $\delta\Omega=2\mu[{1-(\Gamma/(2\mu))^2}]^{1/2}$, where $\Gamma$ is the damping rate of each of the acoustic modes, as shown in Fig. \ref{fig:device}(i). When the acoustic coupling rate is small compared to the dissipation rate ($\mu < \Gamma/2$), the two peaks cannot be resolved, yielding a single-peaked line shape. However, the sharp frequency roll-off is retained, yielding a band-pass filter with a superior shape-factor compared to a typical acoustic Lorentzain response. 

The line shape of the acoustic multi-pole frequency response can be tailored through the device geometry. The resonant frequency of the acoustic modes is readily tunable, demonstrated in Fig. \ref{fig:device}(g), showing the inverse relation between the defect regions width $W$ and the measured resonant acoustic frequency ($\Omega_0 \propto W^{-1}$). The coupling of the two acoustic modes is controlled by the number of rows of holes between the two defects. For example, Figs. \ref{fig:device} (h) and \ref{fig:device} (i) show the acoustic response of devices with $N=5$ and $N=3$ rows of holes between two defect regions of width $W = 3.3 \ \mu$m, yielding acoustic coupling rates ($\mu/(2\pi)$) of 1 MHz and 11 MHz respectively. Further discussion about the phononic crystal design, defect modes, and their coupling is presented in the Supplement, Sections \ref{sec:SI-PnC} and \ref{sec:SI-multi_pole}. 

The use of well-established SOI fabrication techniques allows us to leverage the infrastructure for silicon photonics processing, as well as the favorable properties of silicon, to realize reproducible and robust device performance. Fig. \ref{fig:fab_1}(a) shows the response produced by sixteen devices fabricated on an single chip, all revealing a phononic stopband with at least 50 dB suppression with a highly repeatable two-pole frequency response (more details are shown in the Supplement, Section \ref{sec:SI-fab}). Additionally, the performance of the PPER devices does not show deterioration over time. Fig. \ref{fig:fab_1}(b) shows repeated measurements of a single device, spanning a period of eleven months. The line shape is identical in all measurements, showing a full-width at half-maximum (FWHM) of 6.2 MHz at a center frequency of 3.9 GHz. The measurements were all performed at atmosphere pressure, without environmental control and with no active stabilization. The variation of 0.1\% (4 MHz) in the center frequency can be the result of ambient temperature fluctuations, or the amount of optical power on chip at the time of the measurements, which can readily be stabilized. These results show the potential of these silicon-based devices to be good candidates for practical applications and field deployment.

\begin{figure*}[htb!]
    \centering
    \includegraphics[scale=0.77]{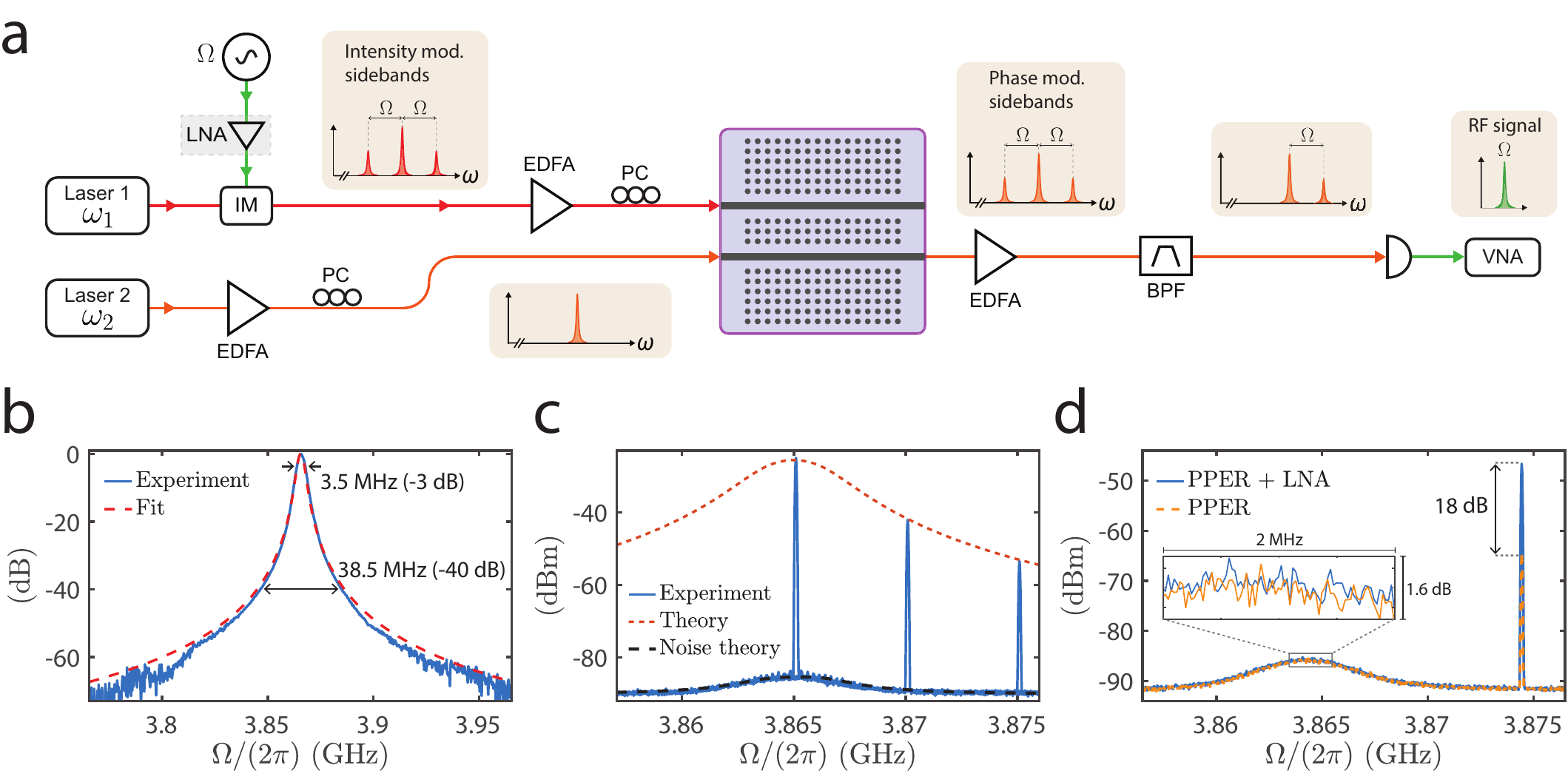}
    \caption{
    \textbf{(a)} Schematic illustration of the experimental setup used to characterise the PPER filter RF-link performance. IM: intensity modulator, LNA: RF low-noise amplifier, EDFA: erbium-doped fiber amplifier, PC: polarization controller, BPF: optical band-pass filter, VNA: RF vector network analyzer.
    \textbf{(b)} Measured pass-band frequency response magnitude, consistent with a two-pole line shape with a FWHM of 3.5 MHz.
    \textbf{(c)} Measurements of single RF tones through the filter, with an input RF-power of $-4$ dBm. The RF frequency was tuned to the center of the pass-band, and at frequencies 5 MHz and 10 MHz from the pass-band center. The two-pole filter response expected from theory \cite{gertler2019microwave} is overlaid  for reference. The thermal-Brillouin noise can also be seen around the filter pass-band center, also in good agreement to theory. The parameters used in the theoretical plots are detailed in the Supplement, Table \ref{tbl:param_table}. The resolution bandwidth (RBW) used in the measurements was 50 kHz.
    \textbf{(d)} Measurements of a single RF tone with RF power $-14$ dBm, at a frequency 10 MHz from the filter center-frequency, with (blue) and without (orange) an RF amplifier at the link input. 
    When using the amplifier, the signal power is 18 dB higher at the filter output, while the noise floor does not change (magnified in inset).
    The bandwidth used in the measurements was 50 kHz.
    }
    \label{fig:filter}
\end{figure*}

\subsection{\label{subsec:filter}Filter performance}
The operation scheme and experimental setup used to test the Brillouin-based filter is illustrated in Fig. \ref{fig:filter}(a). An RF signal at frequency $\Omega$ modulates an optical tone with optical frequency $\omega_1 = 2\pi c/(1.53\ \mu\text{m})$ using an intensity modulator (Optilab IM-1550-20-A), yielding sidebands around the optical carrier. This modulated optical tone is amplified and injected into the `emit' waveguide of the PPER device. When the modulation frequency approaches the Brillouin resonance, acoustic waves are emitted through a forward Brillouin process, resulting in time modulation of the effective refractive index of both waveguides through photo-elastic coupling. A second optical source with frequency $\omega_2 = 2\pi c/(1.53\ \mu\text{m})$ is injected into the `receive' waveguide of the device, where it experiences phase modulation by the transduced acoustic fields. Phase demodulation is implemented by optically filtering one of the sidebands using a commercial band-pass filter (Alnair BVF-300CL), and the signal is detected on a high-power photo-diode (Discovery Semiconductors, Inc. DSC100S, $V_\text{bias}=7$ V). 

First, the frequency response of the filter was measured by sweeping the RF frequency $\Omega$ through the acoustic resonance $\Omega_0$, as shown in Fig. \ref{fig:filter}(b). The filter exhibits a center pass-band frequency of $\Omega_0/(2\pi) = 3.87$ MHz, with a 3-dB linewidth of $\Delta\Omega/(2\pi) = 3.5$ MHz, corresponding to an acoustic \textit{Q}-factor of $\Omega/ \Delta \Omega =  1106$. The second-order filter response shows a fast frequency roll-off of 3.7 dB/MHz, yielding a 40-dB bandwidth of 38.5 MHz, and 70 dB out-of-band suppression at frequencies 100 MHz from the center of the pass-band. In comparison to a Lorentzian line shape with the same FWHM, the two-pole line shape yields a 35 dB improvement of out-of-band suppression. The measured phase response shows a $2\pi$ phase shift over the filter bandwidth as expected from the two-pole filter, corresponding to a group delay of 131 nano-seconds at the center of the pass-band (phase and group-delay data are presented in the Supplement, Section \ref{sec:SI-RF}). The measured two-pole filter response is consistent with a dissipation rate of $\Gamma/(2\pi)= 3.7$ MHz for the acoustic modes, and a coupling rate of $\mu/(2\pi)=1$ MHz. Further discussion of the line shape and fit parameters can be found in the Supplement, Section \ref{sec:SI-multi_pole} and Table \ref{tbl:param_table}.

Next, the properties of the RF link were studied using an RF spectrum analyzer to measure the output RF power as a function of the input RF power. The on-chip optical `emit' power was set to 105 mW and the optical power incident on the detector was 76 mW. The noise floor of the RF link is dominated by thermal-Brillouin fluctuations, a result of the occupation of the acoustic modes at room temperature \cite{kharel2016_Hamiltonian,kittlaus2018rf,gertler2019microwave}. The noise power-spectrum follows a Lorentzian-like line shape with a FWHM of 2.9 MHz, and peak spectral density of $- 134.6 $dBm/Hz. Further details about the noise properties of the RF link are presented in the Supplement, Section \ref{sec:SI-RF}. The RF link had a measured link gain $G = -17.3$ dB, a spurious-free dynamic range $\text{SFDR}_3 = 93.5 $ dB Hz$^{2/3}$, linear dynamic range $\text{CDR}_\text{1dB} = 119.1$ dB Hz, and a noise figure NF = 56.7 dB. The third-order spurious tone is a result of the intensity modulator used at the link input, which can be suppressed using linearization schemes for modulation \cite{khilo2011broadband}. These results can be improved further by using higher optical powers, as well as using a modulator with a lower half-wave voltage \cite{wang2018integrated,gertler2019microwave}. 

The RF-link performance is accurately described by the theoretical analysis presented in Ref. \cite{gertler2019microwave}, as shown in Fig. \ref{fig:filter}(c), using the parameters given in the Supplement, Table \ref{tbl:param_table}. The consistency of the measurements with theory enables a reliable estimation of the RF-link figures of merit from the system parameters, which can be used to design future PPER-based systems. 

The noise figure of the PPER-based RF-photonic link can be improved by increasing the transduction strength in the `emit' waveguide, as the noise floor is set by thermal-Brillouin noise in the `receive' path \cite{gertler2019microwave}. To demonstrate this, an RF amplifier was added before the modulator at the link input (MiniCircuits ZX60-V63+, 18 dB of gain at 4 GHz, noise-figure of 3.8 dB, $V_\text{bias}=5$ V). The addition of the amplifier boosts the signal, without changing the noise floor of the link which is still Brillouin-noise dominated, as shown in Fig. \ref{fig:filter}(d). Using the amplifier, a link gain of $G=0.6$ dB was obtained, with a noise figure of NF = 39.2 dB. We see that the full amount of gain provided by the amplifier (18 dB) contributes directly to a reduction of noise figure. More details about the RF-link measurements are presented in the Supplement, Section \ref{sec:SI-RF} and the measured RF-link parameters are provided in the Supplement, Table \ref{tbl:RF_table}.

\begin{figure*}[ht!]
    \centering
    \includegraphics[scale=0.77]{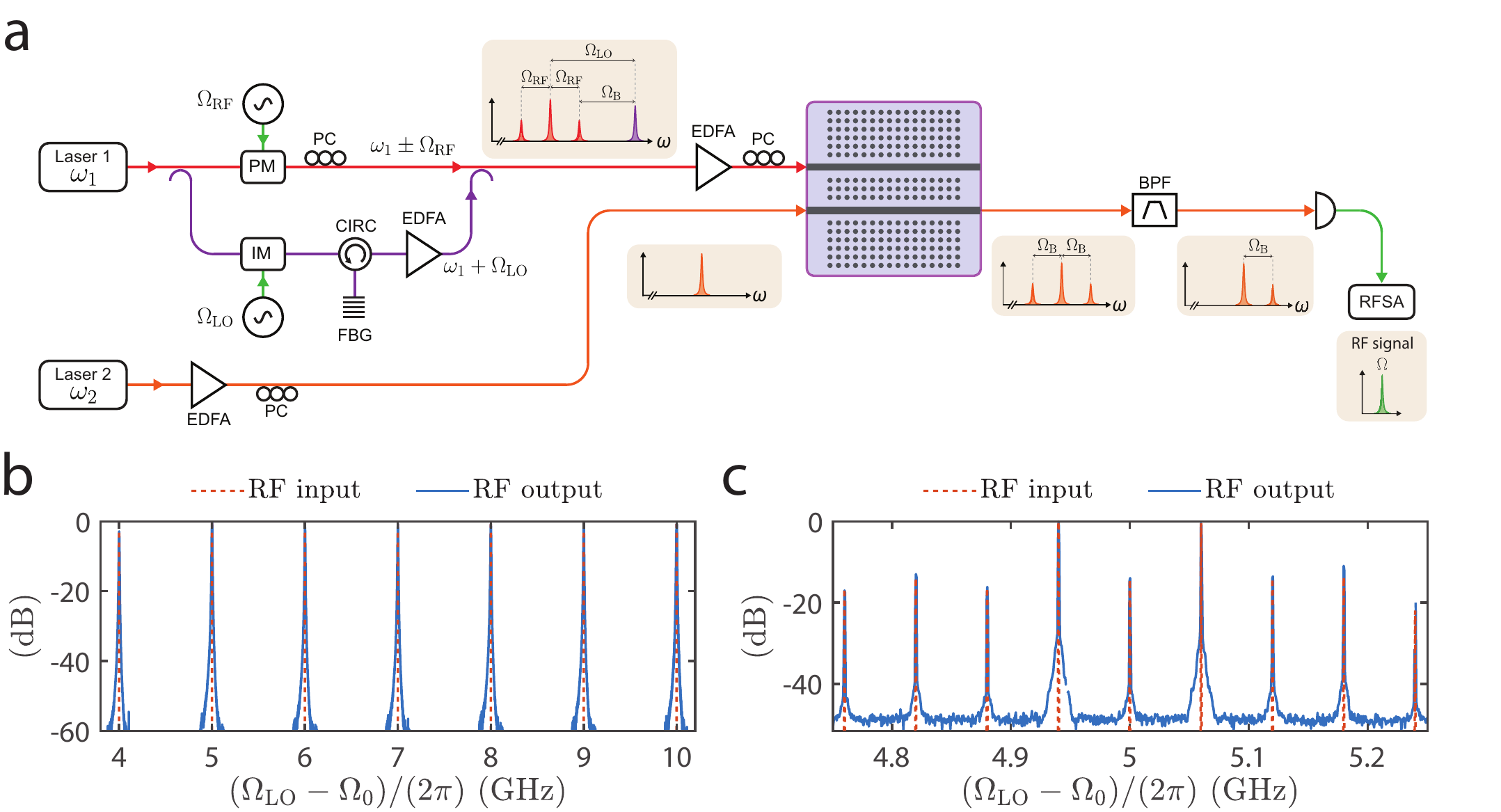}
    \caption{
    \textbf{(a)} Schematic illustration of the experimental setup used to demonstrate filter pass-band tunability. PM: phase modulator, IM: intensity modulator, FBG: fiber-Bragg-grating filter, CIRC: circulator, EDFA: optical amplifier, PC: polarization controller, BPF: optical band-pass filter, RFSA: RF spectrum analyzer.
    \textbf{(b)} Normalized spectrum as a function of local-oscillator frequency, demonstrating the shifting of the filter pass-band over 6 GHz (limited by the phase-modulator bandwidth).
    \textbf{(c)} Normalized spectrum as a function of local-oscillator frequency, with a 60 MHz RF-comb at the filter input. The PPER output reproduces the input, and the resolution is determined by the acoustic line shape.}
    \label{fig:tune}
\end{figure*}
\subsection{\label{subsec:tune}Pass-band tunability}
The pass-band center frequency can be tuned using a modified modulation scheme at the PPER `emit' signal path \cite{gertler2019microwave}. To implement this, the input RF signal is first encoded onto the optical carrier using a phase modulator. Because phonons are generated by intensity modulation of light, this phase-modulated optical carrier alone does not drive a phonon field. To convert the phase-modulated sidebands into an intensity beat-note that can be used to transduce acoustic waves, we introduce a second optical tone, used as an optical local oscillator (LO). The pass-band is set by the Brillouin frequency $\Omega_0$ and the LO offset $\Omega_\text{LO} = \omega_\text{LO}-\omega_1$, which selects the spectral band that is filtered $\Omega_\text{filt} = \Omega_\text{LO}-\Omega_0$. The `receive' path of the PPER is phase modulated at the Brillouin frequency, regardless of the input signal, hence the demodulation and detection at the link output can remain identical to that used in the static filter case.

The experimental setup used to demonstrate the PPER-filter tunability is illustrated in Fig. \ref{fig:tune}(a). The RF input signal was modulated on an optical carrier with optical frequency $\omega_1$ using a phase modulator (Thorlabs LN65S-FC). The optical LO was synthesized from the same laser source, using an intensity modulator and an optical filter, yielding a single tone at optical frequency $\omega_\text{LO} = \omega_1 + \Omega_\text{LO}$. The two optical fields were combined and directed into the `emit' waveguide of the PPER device, while a second optical tone at frequency $\omega_2$ was injected into the `receive' waveguide. The light from the `receive' output was demodulated using a commercial optical filter, detected using a photo-receiver, and measured using an RF spectrum analyzer. 
 
We first demonstrate the wide-band tunability of the filter by shifting the pass-band of the filter ($\Omega_\text{filt}$) over multiple gigahertz. An RF input tone at frequency $\Omega_\text{RF}/(2\pi) = 4$ GHz was injected into the filter input, and the LO swept a 250 MHz range around optical frequency $\omega_\text{LO} = \omega_1 + \Omega_\text{RF} + \Omega_0$. This tunes the filter pass-band $\Omega_\text{filt}=\Omega_\text{LO}-\Omega_0$ around the input tone, showing a sharp peak when $\Omega_\text{filt} = \Omega_\text{RF}$. The input RF tone was shifted by 1 GHz increments up to 10 GHz (limited by the bandwidth of the phase modulator used in the experiment), and the band-pass tuning was repeated at each step. Fig. \ref{fig:tune}(b) presents the aggregated measured traces of the input and output RF signals, showing the pass-band shifting over the 6 GHz range. We see the filter pass-band translated over a GHz range, without changing its narrow-band multi-pole line shape.

To show that the performance of this filter is preserved in the presence of complex RF signals, we synthesized a wideband RF input signal and translated the filter across the spectrum, demonstrating the use of this system for wideband spectral analysis. The input RF signal consisted of a comb of tones separated by 60 MHz, centered around a  5 GHz carrier ($\Omega_\text{RF} = 5$ GHz). By tuning the optical LO, the filter pass-band $\Omega_\text{filt}=\Omega_\text{LO}-\Omega_0$ was swept over a range of 500 MHz around $\Omega_\text{RF}$, as seen in Fig. \ref{fig:tune}(c). In this scenario, the system performs as an RF spectrum analyzer, where the measured RF output reproduces the RF comb at the filter input, with the spectral resolution given by the 6.2 MHz linewidth of the acoustic mode used in the filtering process. The noise floor of this measurement ($\text{SNR}=50$ dB) was determined by the 1 MHz measurement bandwidth. More details about tuning schemes of the PPER RF-photonic filter are presented in the Supplement, Section \ref{sec:SI-tune}.

\section{\label{sec:discussion}Discussion}
In this work, we have demonstrated a tunable RF photonic multi-pole filter in a silicon platform using a PPER scheme. The strong forward Brillouin interaction in the device is used to transduce a microwave signal between the optical and acoustic domains, and in the process, only a narrow spectral band determined by the acoustic response of the device is transduced efficiently. The multi-pole response is a result of the interference between two acoustic modes, yielding sharp frequency roll-off and high out-of-band suppression. This is achieved through acoustic mode-engineering, utilizing acoustic defect modes in a phononic crystal, and coupling acoustic modes in a controllable fashion.

The devices demonstrated in this work were all fabricated using standard SOI wafers and CMOS compatible fabrication methods. While fabrication was performed using electron-beam lithography, identical performance can readily be realized with photo lithography, as the smallest necessary feature size can be 150 nm, readily achievable in CMOS facilities. This can enable the scaling of production of silicon-PPER devices, yielding cheap, high-volume, and consistent results \cite{thomson2016roadmap,chen2018emergence}.

The RF-link performance using the silicon PPER device is competitive with other RF-photonic schemes \cite{marpaung2019integrated} such as ring-resonators \cite{liu2017all}, interferometers \cite{fandino2017monolithic}, Brillouin-based filtering \cite{eggleton2019brillouin}, and pulse shaping \cite{metcalf2016integrated}. Furthermore, the line shape obtained in the PPER scheme stands out for its ability to produce multi-pole frequency responses with a narrow-band (3.5 MHz), excellent out-of-band suppression (70 dB) and a wide stop-band (2 GHz). 
Similar performance in the optical domain using ring resonators would require two ultrahigh-$Q$ ($>10^8$) resonators, and ultralow-loss ($\sim0.1$ dB/m) waveguides \cite{little2004very,dong2010ghz,tien2011ultra,onural2020ultra}. Implementing such a device in silicon would demand a large footprint \cite{biberman2012ultralow}, as well as sub-millikelvin temperature stability \cite{jayatilleka2015wavelength} and narrow-linewidth laser sources. 
In contrast, the devices in this work have a footprint of $\sim0.1$ mm$^2$, which can facilitate integration for filter-bank and channelizing applications. The measurements presented here were performed with no active stabilization, as the resonant frequencies are determine in the acoustic domain. In order to avoid drift of the filter pass-band over longer periods of time, the temperature needs to be stabilized on the order of $\sim1$ K. 
Additionally, the PPER design can be extended to produce higher order filters \cite{gertler2019shaping}. As an example, we present third-order filters in the Supplement, Section \ref{subsec:SI-multi_pole-3_pole}.

The PPER-based filtering scheme utilizes the optical, microwave, and acoustic domains --- each with vastly different and complementary properties --- which can all be optimized for further improvement of the device performance. Longer devices will result in higher gain, lower noise figure, and a larger dynamic range \cite{gertler2019microwave}. For example, by using a PPER device with an active length of 4 cm the link gain will increase by 20 dB, while the noise figure will reduce by 10 dB compared to the results shown in this work, which were obtained from a 4 mm long device \cite{kittlaus2018rf,gertler2019microwave}. Higher optical power in the `emit' waveguide will yield a higher link gain, without adding noise in the process. Dispersion engineering of the optical waveguides can enable stronger Brillouin coupling \cite{merklein2015enhancing}, resulting in a larger spur-free dynamic range and lower noise figure. Additionally, an interferometric phase demodulation scheme will result in a higher link gain \cite{urick2015fundamentals,gertler2019microwave}. In the acoustic domain, stronger acousto-optic coupling can be achieved by longer lived acoustic modes, which will also result in a narrower filter line shape. For example, this can be accomplished by optimizing the phononic-crystal design, achieving a full acoustic band-gap \cite{mohammadi2008evidence}. In the microwave domain, the introduction of improved electro-optic modulators will directly enhance the performance of the link. In the scheme we have demonstrated, the third-order spurious tone is a result of the intensity modulator at the filter input, and using linearized modulation schemes suppressing these spurious tones \cite{khilo2011broadband} will yield a larger dynamic range. Further, modulators with lower half-wave voltage \cite{wang2018integrated} will result in a lower noise figure \cite{gertler2019microwave}. Alternatively, a microwave low-noise amplifier (LNA) can be used at the input to the modulator, yielding a lower noise figure, as was demonstrated in this work. 

We have shown how the filter pass-band can be tuned over a range of multiple GHz, by using an optical tone as a local oscillator.
The bandwidth limitation of this scheme lies in the dual-sideband nature of the forward Brillouin process \cite{kang2009tightly,kittlaus2016large,kharel2016_Hamiltonian}, similar to the distortion from an image frequency in heterodyne receivers \cite{ mirabbasi2000classical}. This restricts the bandwidth of the input RF signal to be smaller than twice the Brillouin frequency for distortion-free operation (for more details see Supplement, Section \ref{subsec:SI-tune-bandwidth}). This bandwidth limitation can be avoided by utilizing a single-sideband process to generate the acoustic field, such as inter-modal Brillouin scattering \cite{kittlaus2017_Intermodal,kittlaus2018non}. In our demonstration of filter tuning, the RF signal was shifted to the Brillouin frequency at the filter output, beneficial in systems where both filtering and frequency conversion are required \cite{mirabbasi2000classical}. However, by designing an alternative demodulation scheme, a frequency-preserving RF-photonic filter can be achieved (for further discussion, see Supplement, Section \ref{sec:SI-tune}).

The forward Brillouin process used in the filtering scheme demonstrated here has the advantage of scalability by cascading in devices in series. Multiple PPER devices could be integrated on the same chip, without losing signal fidelity as the input signal traverses the device array \cite{gertler2019shaping}. This can enable channelizing and sensing schemes with superior performance compared to systems where the signal is split and amplified in multiple stages, adding noise in the process. The high signal-to-noise of the PPER-based link enables the use of short segments of active-Brillouin regions, resulting in a small overall footprint \cite{kittlaus2018rf,gertler2019microwave}. 

With the advancement of integrated silicon light sources \cite{rong2005continuous,otterstrom2018silicon}, amplifiers \cite{foster2006broad,kittlaus2018rf}, modulators \cite{reed2010silicon} and detectors \cite{baehr2005optical}, PPER-based filtering schemes can be an important step towards foundry-compatible, fully integrated RF-photonic systems.

\section{\label{sec:ack}Acknowledgments}
We thank Taekwan Yoon, Prashanta Kharel, and David Mason for discussions involving RF-link measurements, phononic-phononic systems, and nonlinear interactions.
This research was supported primarily by the Packard Fellowship for Science and Engineering.
We also acknowledge funding support from ONR YIP (N00014-17-1-2514).
N.T.O acknowledges support from the National Science Foundation Graduate Research Fellowship under Grant DGE1122492.
Part of the research was carried out at the Jet Propulsion Laboratory, California Institute of Technology, under a contract with the National Aeronautics and Space Administration.

\onecolumngrid
\newpage
\appendix
\maketitle
\tableofcontents

\newpage
\section{\label{sec:SI-fab}Device fabrication}
The devices presented in this work were fabricated on a single-crystal silicon-on-insulator (SOI) chip with a layer structure of 215 nm silicon on a 3 $\mu$m oxide layer. A first electron-beam lithography step defines the optical waveguide rib structures using hydrogen silsesquioxane (HSQ) electron-beam resist and development in MF-312 and a Cl$_2$ reactive ion etch (RIE), removing 80 nm of silicon. A second lithography step is used to define the array of holes with CSAR electron-beam resist, followed by development in Xylenes. The remainder of the silicon is removed through another Cl$_2$ RIE, exposing the oxide. Finally, a wet-etch using 49\% hydrofluoric (HF) acid removes the oxide under-cladding to create suspended Brillouin-active waveguides. Light was coupled on and off the chip using integrated grating couplers.

The Brillouin-active region of the device demonstrating the RF link, shown in Fig. \ref{fig:filter}, was $L=5$ mm long, with $N=5$ lines of holes between the two waveguides, and phononic defect regions of width $W=3.3 \ \mu$m. The phononic-crystal pitch was $a=650$ nm with a hole diameter of $d=500$ nm.
The device used to demonstrate the filter tunability, shown in Fig. \ref{fig:tune}, had a Brillouin-active region length of $L=3$ mm, and defect regions width $W=3.3 \ \mu$m spaced by $N=5$ lines of holes between the two waveguides. The phononic-crystal pitch was $a=600$ nm and the holes diameter $d=462$ nm. The resonant acoustic frequency was $\Omega_0=3.896$ GHz, with a FWHM of 6.2 MHz.

\begin{figure}[hb!]
    \centering
    \includegraphics[scale=0.77]{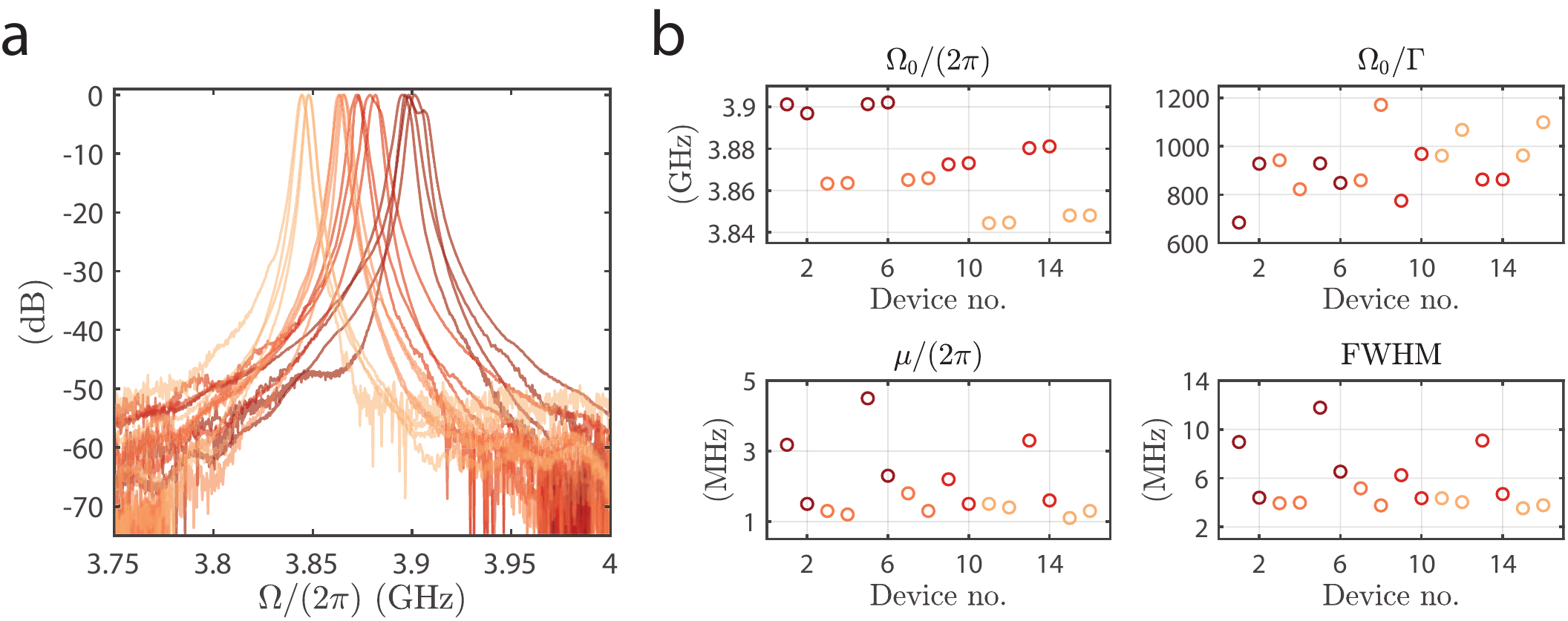}
    \caption{
    \textbf{(a)} Measurements of sixteen devices fabricated on the same chip, all showing the two-pole frequency response from the phonon defect modes around 3.855 GHz, corresponding to the larger span presented in Fig. \ref{fig:fab_1}(a) in the main text.
    The variation of the center frequency and precise line shapes are the result of a slightly different geometry of each of the devices.
    \textbf{(b)} Fitted values to the filter line shapes, showing high consistency in the center frequency ($\Omega_0$), $Q$-factor of the acoustic modes ($\Omega_0/\Gamma$), acoustic coupling rate ($\mu$), and full-width at half-maximum (FWHM) of the two-pole line shape.
    }
    \label{fig:fab_2}
\end{figure}

Fig. \ref{fig:fab_2}(a) demonstrates the consistency and robustness of the fabrication process, by showing measurements of 16 devices fabricated on the same chip, with 100\% yield in the fabrication process. All the devices show good performance, with an acoustic stop-band of 1.5 GHz with 50 dB suppression, and a defect modes in the range 3.85--3.9 GHz. Examining the filter shapes more closely reveals that all of the devices produce a two-pole line shape with FWHM of a few MHz, fitted parameters are displayed in Fig. \ref{fig:fab_2}(b). The line shapes vary slightly between devices, consistent with different design parameters in the geometry of each device such as the number of hole rows between the defect modes, and the defect region width. The performance of the devices and the repeatability of the fabrication can be further improved by using photo-lithography in a dedicated SOI foundry, which can yield higher consistency and reduce alignment and drift issues present in the electron-beam lithography used in this work.

The fabrication steps used in this work are mostly typical CMOS lithography techniques. The last step in the process, where the waveguides are suspended, is commonly used in MEMS fabrication, and is only necessary around the active-Brillouin regions of the device. This could be performed post-process, or incorporated into the fabrication procedure. The dimensions of the smallest features defined in the lithography were on the order of $\sim$150 nm.

\clearpage
\section{\label{sec:SI-PnC}Phononic crystal design}
\begin{figure*}[hb!]
    \centering
    \includegraphics[scale=0.77]{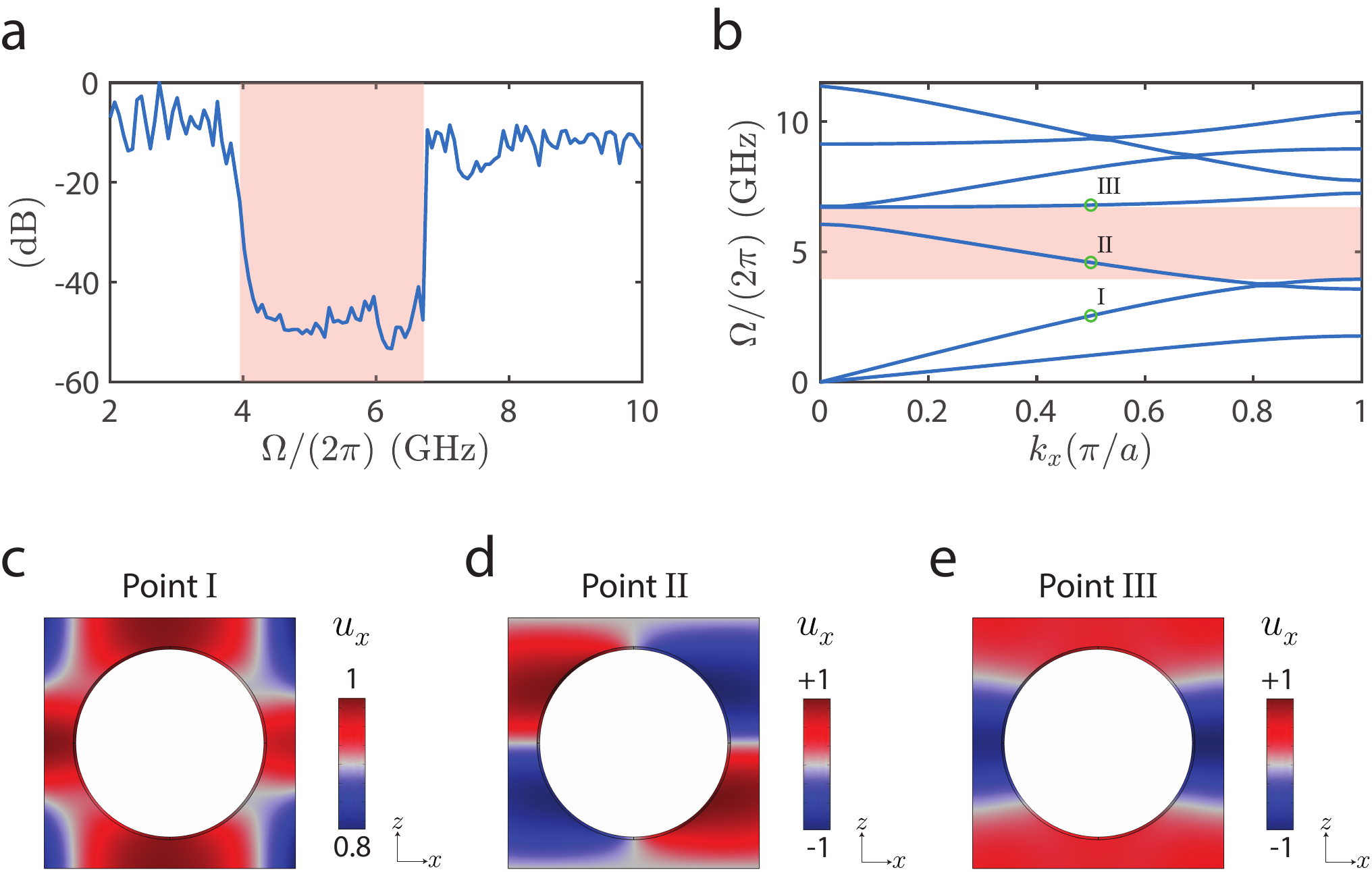}
    \caption{
    \textbf{(a)} Simulated acoustic transmission of an acoustic wave with $x$ displacement trough twelve rows of a phononic crystal, showing a 2.5 GHz stop-band. The cubic phononic crystal in the simulation has a pitch of $a = 650$ nm and hole diameter of $d = 500$ nm.
    \textbf{(b)} Simulated 2D band structure of the phononic crystal, along the $\Gamma-\text{X}$ line of the Brillouin zone. The modes in the stop-band do not couple strongly to the optical fields.
    \textbf{(c-e)} The $x$-component of the displacement $u_x$ of a unit cell in the cubic lattice at points \MakeUppercase{\romannumeral 1}, \MakeUppercase{\romannumeral 2} and \MakeUppercase{\romannumeral 3} from (b), respectively.
    }
    \label{fig:PnC}
\end{figure*}

The cubic lattice of air holes in the silicon structure results in a stop-band for the Brillouin-active acoustic modes guided in the devices. This is demonstrated using a finite-element-method (FEM) simulation, presented in Fig. \ref{fig:PnC}(a), showing low acoustic transmission in the range between 3.9 and 6.7 GHz. The simulated cubic lattice has 12 rows of holes with a pitch of $a = 650$ nm and hole diameter of $d = 500$ nm. 
A simulation of the acoustic eigen-modes of a single unit cell yields the acoustic band structure presented in Fig. \ref{fig:PnC}(b) for wave-vectors along the $k_x$ axis. As seen from the band structure, there is not a complete band gap across this frequency region. An analysis of the bands shows that the acoustic modes in the stop-band do not couple to the optical fields through the forward Brillouin interaction utilized in the device.
Figs. \ref{fig:PnC}(c-e) present the x-component of the displacement of the modes at wave-number $k_x = \pi/(2a)$, showing even and odd symmetries for the modes outside and inside the stop-band, respectively.
The acoustic mode used in the forward Brillouin interaction in the device are uniform along the $z$ axis \cite{shin2015control,gertler2019shaping}, hence will not couple strongly to the odd-symmetric mode shown in Fig. \ref{fig:PnC}(d), resulting in the acoustic stop-band.
The difference in frequency obtained in the simulation when compared to the measured stop-bands, could be a result of the different geometry of the fabricated devices, which include the rib-waveguide structure, as well as built-in stress in the silicon layer, which is a result of the undercut of the oxide layer.

\clearpage
\section{\label{sec:SI-geom}Frequency-response trimming through device geometry}
\begin{figure*}[hb!]
    \centering
    \includegraphics[scale=0.77]{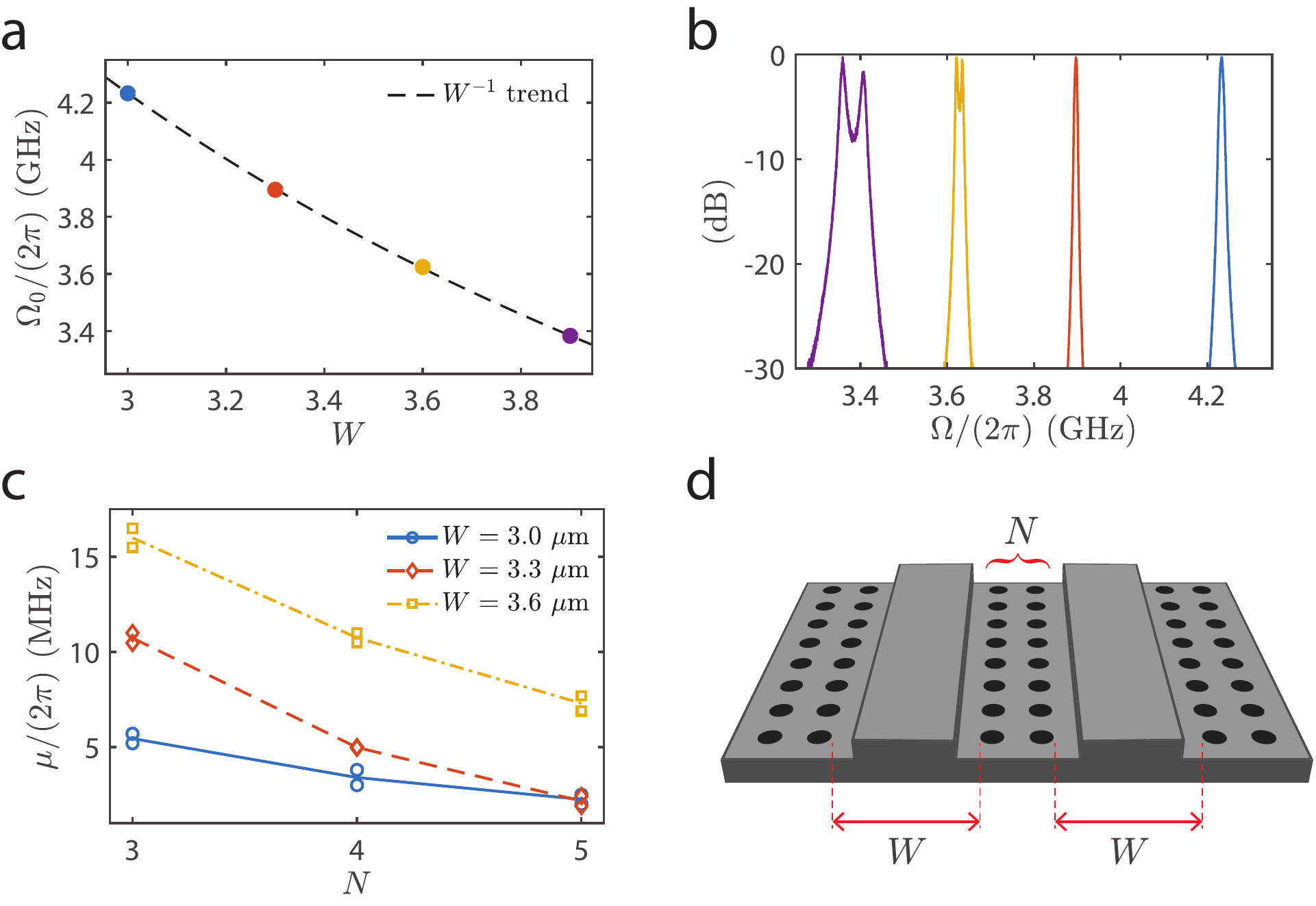}
    \caption{
    \textbf{(a)} The measured resonant frequencies of devices fabricated with different line-defect widths follow an inverse trend $\Omega_0 \propto W^{-1}$, enabling a high degree of control of the filter pass-band through the device design.
    \textbf{(b)} Normalized measured frequency response corresponding to the data-points in panel (a), showing the shift in the acoustic resonance. The difference in the filter line-shape is a result of the different coupling rates for different acoustic frequencies. The phononic crystal design used in these devices has a pitch of $a = 600$ nm and hole diameter of $d = 462$ nm.
    \textbf{(c)} Fitted coupling rates ($\mu$) for measured two-poles devices, with different numbers of rows of holes ($N$). The two data points at each value of $N$ correspond to two different devices.
    \textbf{(d)} A schematic illustration of a two-pole PPER device. The length $W$ is the width of each line-defect in the phononic crystal, and $N$ denotes the number of rows of holes between the two defect regions.
    }
    \label{fig:geom}
\end{figure*}

The long-lived (high-$Q$) acoustic modes utilized in the PPER operation are produced using defects introduced to the phononic crystal lattice. These are in the form of missing rows of holes, as shown in \ref{fig:geom}(d), where a section of the device is illustrated schematically, with defects of width $W$. The number of hole rows between the defect regions is denoted $N$. The width of the defect region determines the frequency of the acoustic resonance which can be utilized for strong forward Brillouin coupling \cite{shin2013tailorable}.
By fabricating devices with defect-region widths between 3 and 3.9 $\mu$m, the measured resonant frequency of the PPER operation was in the range of 3.38 and 4.23 GHz, as shown in Fig. \ref{fig:geom}(b). The frequency follows an inverse relation to the width ($\Omega_0 \propto W^{-1}$), as shown in Fig. \ref{fig:geom}(a), demonstrating how the filter frequency can be tuned through the design of the device geometry.

The two acoustic resonances are coupled in the region between the two line defects, consisting of $N$ rows of holes. By fitting the measured frequency responses, we can estimate the coupling rate $\mu$ between the two modes, further discussed in Section \ref{sec:SI-multi_pole}. As shown in Fig. \ref{fig:geom}(c), the coupling rate drops as hole rows are added, consistent with the decay of the acoustic field in the phononic crystal stop-band.
We can also see that the coupling rate is higher for wider defect modes. These correspond to lower frequency modes, having longer wavelengths, such that effectively the distance between the two line defects is shorter. This is also visible in the measurements shown in \ref{fig:geom}(b), where the lower frequency resonances show a frequency splitting corresponding to a high coupling rate of the two acoustic modes. The dependence of the frequency response on the acoustic coupling rate is further discussed in Section \ref{sec:SI-multi_pole}.

\clearpage
\section{\label{sec:SI-RF}RF-photonic link}
\subsection*{\label{subsec:SI-RF-freq}Frequency response}
\begin{figure*}[htb!]
    \centering
    \includegraphics[scale=0.77]{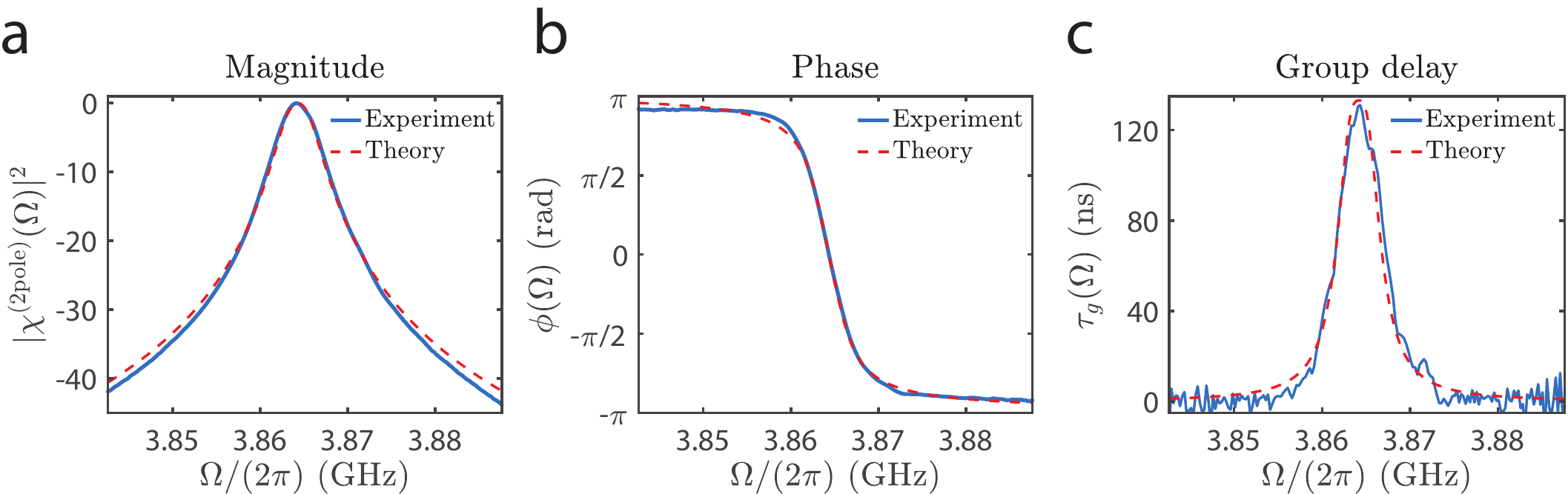}
    \caption{
    \textbf{(a)} Normalized magnitude of the filter frequency response, following a two-pole line shape with a FWHM of 3.5 MHz. 
    \textbf{(b)} The measured phase response shows a $2\pi$ phase-shift over the filter bandwidth. An overall linear term has been removed in the analysis.
    \textbf{(c)} Measured group delay of the filter, showing a maximum of $\tau_g=131$ ns at the center of the filter pass-band.
    }
    \label{fig:VNA}
\end{figure*}

We characterize the filter frequency response using the experimental setup illustrated in Fig. \ref{fig:filter}(a) in the main text. 
The magnitude of the frequency response shows a two-pole line shape, with a center frequency at 3.87 GHz and a full-width at half-maximum (FWHM) of 3.5 MHz.
The two-pole frequency response can be described using \cite{gertler2019shaping}
\begin{equation}
    \chi^\text{(2 pole)}(\Omega) \propto \frac{-i \mu}{ \left[\Omega-\left(\Omega_0+\mu- i\Gamma/2\right) \right] \left[\Omega-\left(\Omega_0-\mu- i\Gamma/2\right) \right]},
    \label{eq:2pole_cmplx}
\end{equation}
which is determined by the acoustic dissipation rate $\Gamma$ and the acoustic coupling rate $\mu$. Fitting Eq. (\ref{eq:2pole_cmplx}) to the data yields parameter values of $\Gamma/(2\pi) = 3.7$ MHz and $\mu/(2\pi) = 1$ MHz, as shown in Fig. \ref{fig:VNA}(a).
Fig. \ref{fig:VNA}(b) presents a phase measurement of the second-order frequency response, showing an overall $2\pi$ phase shift across the filter, consistent with Eq. (\ref{eq:2pole_cmplx}). 
The group delay of the filter, corresponding to the time it takes for an amplitude envelope to propagate through the link, was measured at $\tau_g=131$ ns at the center of the pass-band, as shown in Fig. \ref{fig:VNA}(c). 
This is in good consistency with theory, where the group delay is given by the derivative of the phase response
\begin{equation}
    \phi(\Omega) = \arg\left( \chi^\text{(2 pole)}(\Omega) \right), \qquad \tau_g (\Omega)= -\frac{d \phi (\Omega')}{d \Omega'} \biggr\rvert_{\Omega'=\Omega}.
    \label{eq:group_delay}
\end{equation}

\subsection*{\label{subsec:SI-RF-noise}Thermal-Brillouin noise}
\begin{figure*}[hb!]
    \centering
    \includegraphics[scale=0.77]{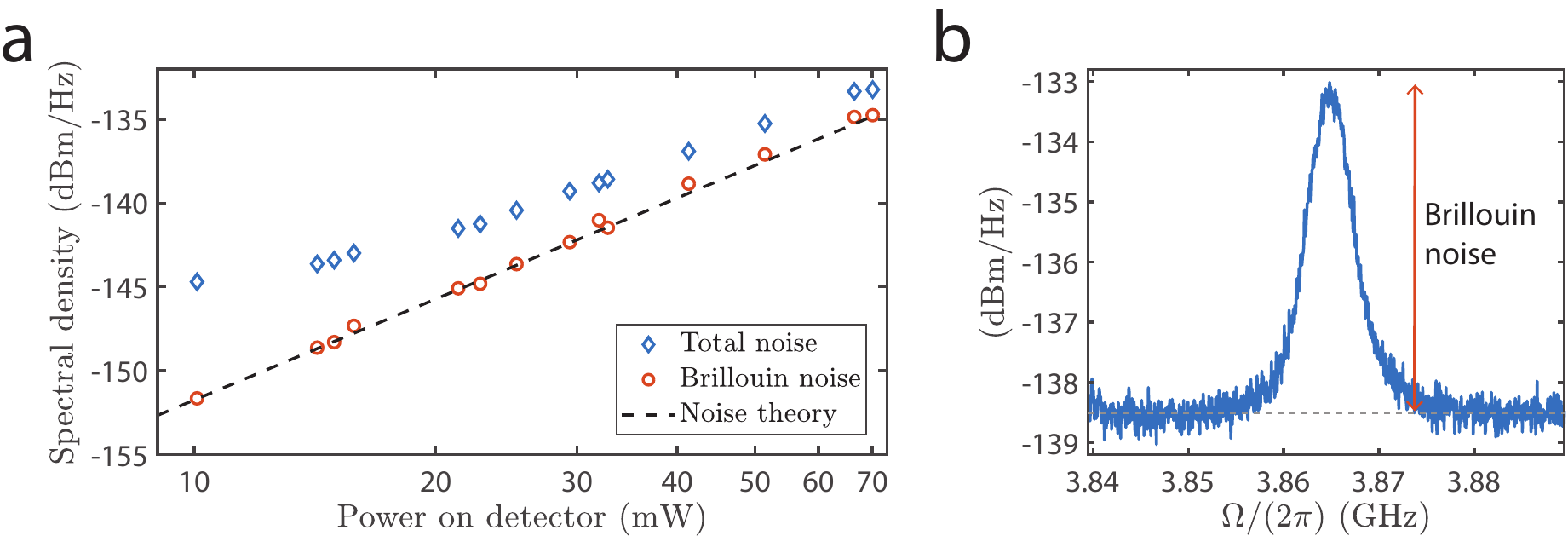}
    \caption{
    \textbf{(a)} Measured peak-value of noise power spectral-density, as a function of the optical power on detector (blue). The extracted thermal-Brillouin contribution to the noise (red) fits the trends predicted from the theory (Eq. (\ref{eq:RF_link_theory})). 
    \textbf{(b)} An example  of the measured noise spectral density, when the optical power on the detector is 70 mW, showing the narrow-band thermal Brillouin peak.
    }
    \label{fig:noise}
\end{figure*}

%
The noise floor of the RF link is comprised of a Lorentzian-like peak with a FWHM of 2.9 MHz, a result of spontaneous Brillouin scattering, and a wide-band noise background (dominated by EDFA noise), shown in Fig. {\ref{fig:noise}}(b). 
The spontaneous Brillouin scattering is a result of the thermal occupation of the phonon modes at room temperature, given by $k_\text{B} T  /(\hbar \Omega_0) = 1565$.
Fitting the Brillouin contribution to the noise power spectrum shows good agreement with the theoretical values expected for the system parameters \cite{gertler2019microwave}, as seen in Fig. {\ref{fig:noise}}(a). 
It is important to note that with our system parameters, the noise floor is not dependent on the RF input nor on the optical field in the `emit' waveguide, and is fully determined by the `receive' path parameters \cite{kittlaus2018rf,gertler2019microwave}. The calculated RF-link gain and noise power-spectral density (per bandwidth $B_\text{RF}$) are given by \cite{gertler2019microwave}
\begin{equation}
    \begin{split}
    \frac{P^\text{RF}_\text{out}(\Omega)}{P_\text{in}^{\text{RF}}} &= \frac{1}{4} R_\text{in} \left( \frac{\pi}{V_{\pi}} \right)^2 \left( P^\text{(rec)} \sqrt{G^\text{(rec)}_\text{B} G^\text{(emit)}_\text{B}} P^\text{(emit)} L\right)^2 \left(\frac{\Gamma}{2} \left|\chi^\text{(2pole)} \right|\right)^2 \eta^2 R_\text{out} |H_\text{pd}|^2,\\
    \frac{P_\text{N}(\Omega)}{B_\text{RF}} &= 2 \frac{\omega_0}{\Omega_0} \Big(k_\text{B} T \Big) \left( G^\text{(rec)}_\text{B} {P^\text{(rec)}}^2 L \right) \left(\frac{\Gamma}{2} \left|\chi^\text{(rec)}_N\right|\right)^2 \eta^2 R_\text{out} |H_\text{pd}|^2 + N_\text{bg},
    \end{split}
    \label{eq:RF_link_theory}
\end{equation}
where $\eta$ is the detector responsivity, $R_\text{out}$ the detector impedance, $H_\text{pd}$ the photodiode circuit efficiency,
$\omega_0$ the `receive' optical frequency, and $k_\text{B}$ denotes the Boltzmann constant. The frequency responses $|\chi^\text{(2pole)}|$ and $|\chi^\text{(rec)}_N|$ are defined in Section \ref{sec:SI-multi_pole}, Eqs. (\ref{eq:2pole_mag}) and (\ref{eq:2pole_N}), and a description of the rest of the parameters can be found in Table \ref{tbl:param_table}.

The first term in the expression for the noise power-spectral density is the thermal-Brilloin noise, and $N_\text{bg}$ accounts for the wide-band noise background.
Theoretical predicted values for the signal and the noise show good agreement with the measurements, as can be seen in Fig. \ref{fig:filter}(c) of the main text (parameter values are presented in Table \ref{tbl:param_table}).
\begin{table*}[!htbp]
\centering
\setlength{\tabcolsep}{8pt} 
\renewcommand{\arraystretch}{1.4} 
\begin{tabular}{| l | c | l |} \hline
Parameter & Value & Description  \\ \hline\hline
$\lambda$ (nm) & 1530 & Optical wavelength \\ \hline
$T$ (K) &  290 & Temperature  \\ \hline
$\Omega_0$ (2$\pi$ GHz) & 3.86 & Phonon frequency \\ \hline
$\Gamma$ (2$\pi$ MHz) & 3.7 & Acoustic dissipation rate \\ \hline
$\mu$ (2$\pi$ MHz) & 1 & Acoustic coupling rate \\ \hline
$V_+^\text{(A)}$ & $\sqrt{0.48}$ & Asymmetry parameter \footnote{Defined in Section \ref{sec:SI-asym}.}\\ \hline
$P^\text{(emit)}$ (mW) & 113 & Optical power in the `emit' waveguide \\ \hline
$P^\text{(rec)}$ (mW) & 76 & Optical power on the detector \\ \hline
$B_\text{RF}$ (kHz) & 50 & RF-measurement bandwidth (RBW) \\ \hline
$G^\text{(emit)}_\text{B}$ ($\text{W}^{-1}\text{m}^{-1}$) & 1300 & Brillouin gain in `emit' waveguide \\ \hline
$G^\text{(rec)}_\text{B}$ ($\text{W}^{-1}\text{m}^{-1}$) & 915 & Brillouin gain in `receive' waveguide \\ \hline
$L$ (mm) & 5 & Active-Brillouin interaction length \\ \hline
$V_\pi$ (V) & 6.94 & Half-wave modulation voltage \footnote{The modulator was not biased at the quadrature point.} \\ \hline
$\eta^2 R_\text{out} |H_\text{pd}|^2$ ($\text{W}^{-1}$)& 4.6 & Calibrated detector response \\ \hline
$R_\text{in}$ ($\Omega$) & 50 & Intensity modulator input impedance \\ \hline
$P_\text{in}^\text{RF}$ (dBm) & 10.1 & Input RF power \\ \hline
$N_\text{bg}$ (dBm/Hz) & $-139.1$ & Background noise \\ \hline
\end{tabular}
\caption{Parameters used for the theory plots in Figs. \ref{fig:filter}(c) and \ref{fig:noise}(a).}
\label{tbl:param_table}
\end{table*}

\subsection*{\label{subsec:SI-RF-link}RF-link properties}
\begin{figure*}[hb!]
    \centering
    \includegraphics[scale=0.77]{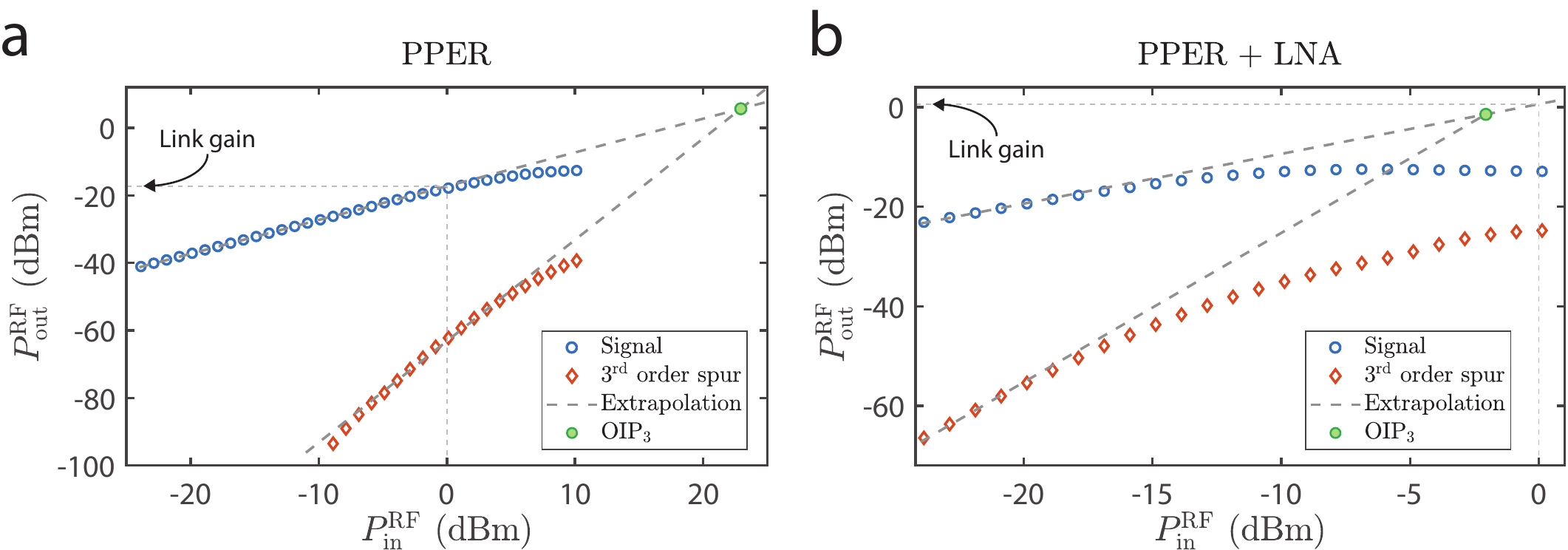}
    \caption{
    \textbf{(a)} Measured RF output power as a function of RF input power for the fundamental (blue) and third harmonic (red) tones of the filter. Extrapolated linear trends and the third-order intercept point (OIP$_3$) and link-gain are shown for reference. On-chip `emit' power was 105 mW and the power incident on the detector was 76 mW. The bandwidth used in the measurement was 300 Hz.
    \textbf{(b)} Repeated measurement when cascading an RF amplifier at the filter input. 
    The RF-link parameters extracted from the measurements are presented in Table \ref{tbl:RF_table}.
    }
    \label{fig:RF}
\end{figure*}

The RF link is characterized by measuring the RF power at the output of the filter ($P^\text{RF}_\text{out}$) as a function of the input RF power ($P^\text{RF}_\text{in}$). Fig. \ref{fig:RF}(a) shows the link performance, with incident on-chip optical power of 105 mW, and 76 mW of `receive' optical power on the detector. The measured small-signal gain is $G=-17.3$ dB and the noise floor $N = -134.6$ dBm/Hz. Assuming room temperature thermal noise as the input noise source ($N_\text{in} = k_\text{B}T = -174$ dBm/Hz) yields a noise-figure of $\text{NF} = N - G - N_\text{in} = 56.7$ dB. The 1 dB compression point was measured at $P_\text{in}^\text{1dB}= 1.7$ dBm, yielding a linear dynamic range of $\text{CDR}_\text{1dB} =P_\text{in}^\text{1dB}+G-N= 119.1$ dB Hz. We next quantify the spurious-free dynamic range by injecting an RF tone at frequency $\Omega = \Omega_0/3$, and measuring the output RF power at frequency $\Omega_0$, shown in Fig. \ref{fig:RF}(a). The extrapolated third-order intercept point was $\text{IIP}_3 = 22.9$ dBm, yielding a spurious-free dynamic range of $\text{SFDR}_3 = (2/3)(\text{IIP}_3 + G - N) =93.5$ dBm Hz$^{2/3}$. The measured $\text{IIP}_3$ of the RF link is mainly deterined by the linearity of the intensity modulator at the filter input (a direct measurement of the modulator alone yielded an intercept point of $\text{IIP}_3^\text{IM} = 24.3$ dBm). This suggests that the spurious-free dynamic range can be directly improved by using a linearized modulation scheme at the link input \cite{khilo2011broadband}.

The noise-figure of the PPER-based RF link can be further improved by using an RF amplifier (LNA) at the link input (front-end). This is a result of thermal-Brillouin scattering being the dominant noise source, such that the added noise from the LNA is negligible at the link output \cite{gertler2019microwave}. We demonstrate this by adding an LNA (MiniCircuits ZX60-V63+, 18 dB of gain at 4 GHz, noise-figure of 3.8 dB, $V_\text{bias}=5$ V) at the link input (Fig. \ref{fig:filter}(a)). 
The input RF tone was tuned 10 MHz off the center of the pass-band in order to clearly show the thermal-Brillouin noise floor at the pass-band center. 
Fig. \ref{fig:filter}(d) in the main text superimposes two spectrum measurements, with and without the LNA, showing how the noise floor is unaffected by the presence of the amplifier.
In contrast, the power of the output RF signal is amplified by 18 dB when the LNA is added, demonstrating how this scheme yields higher signal power without a noise penalty, reducing the noise-figure by the amount of gain provided by the LNA.

We repeat the RF power measurements with the LNA at the link input, as presented in Fig. \ref{fig:RF}(b). The measured link gain is now $G = 0.6$ dB, while the output noise is unchanged, measured at $N = -134.2$ dBm/Hz, and resulting in a noise-figure of $\text{NF} = 39.2$ dB. We see an improvement of 17.5 dB, corresponding to the gain of the LNA. The linear dynamic range is $\text{CDR}_\text{1dB} = 113.6$ dB Hz, unchanged by the LNA, as expected from a link analysis presented in Ref. \cite{gertler2019microwave}. The measured spurious-free dynamic range is $\text{SFDR}_3 = 88.5$ dBm Hz$^{2/3}$, showing the trade-off between noise-figure and dynamic range in this scheme, which is a common situation in many microwave-photonic links \cite{bucholtz2008graphical}. The measured RF-link parameters (with and without the LNA) are summarized in Table \ref{tbl:RF_table}.

\begin{table*}[!htbp]
\centering
\setlength{\tabcolsep}{8pt} 
\renewcommand{\arraystretch}{1.4} 
\begin{tabular}{| l | c | c | l |} \hline
Parameter & PPER & PPER + LNA & Description  \\ \hline\hline
$G$ (dB) & $-17.3$ & 0.6 & RF link gain $\left(P_\text{out} / P_\text{in}\right)$  \\ \hline
$N$ (dBm/Hz) & $-134.6$ & $-134.2$ & Noise floor $\left(P_\text{out} / P_\text{in}\right)$  \\ \hline
$\text{NF}$ (dB) & 56.7 & 39.2 & Noise figure $\left( \text{SNR}_\text{in}/\text{SNR}_\text{out} \right)$ \\ \hline
$\text{OIP}_3$ (dBm) & 5.6 & $-1.5$ & Output intercept point  \\ \hline
$\text{SFDR}_3 \ \big(\text{dB Hz}^{2/3}\big)$ & 93.5 & 88.5 & Spur-free dynamic range $\left(\text{OIP}_3/P_\text{N}\right)^{2/3}$  \\ \hline
$P_\text{in}^{\text{1dB}}$ (dBm) & 1.7 & $-15.2$ & Input 1 dB compression point  \\ \hline
$\text{CDR}_{1\text{dB}}$ (dB Hz) & 119.1 & 119.6 & Linear dynamic range $\left( P_\text{out}^{\text{1dB}} / P_\text{N}\right)$  \\ \hline
\end{tabular}
\caption{Measured RF-link parameters of the two-pole PPER-based RF-photonic filter, with and without an RF amplifier at the link input, corresponding to the data presented in Fig. \ref{fig:RF}}
\label{tbl:RF_table}
\end{table*}

\clearpage
\section{\label{sec:SI-tune}Filter tunability}
\begin{figure*}[hb!]
    \centering
    \includegraphics[scale=0.64]{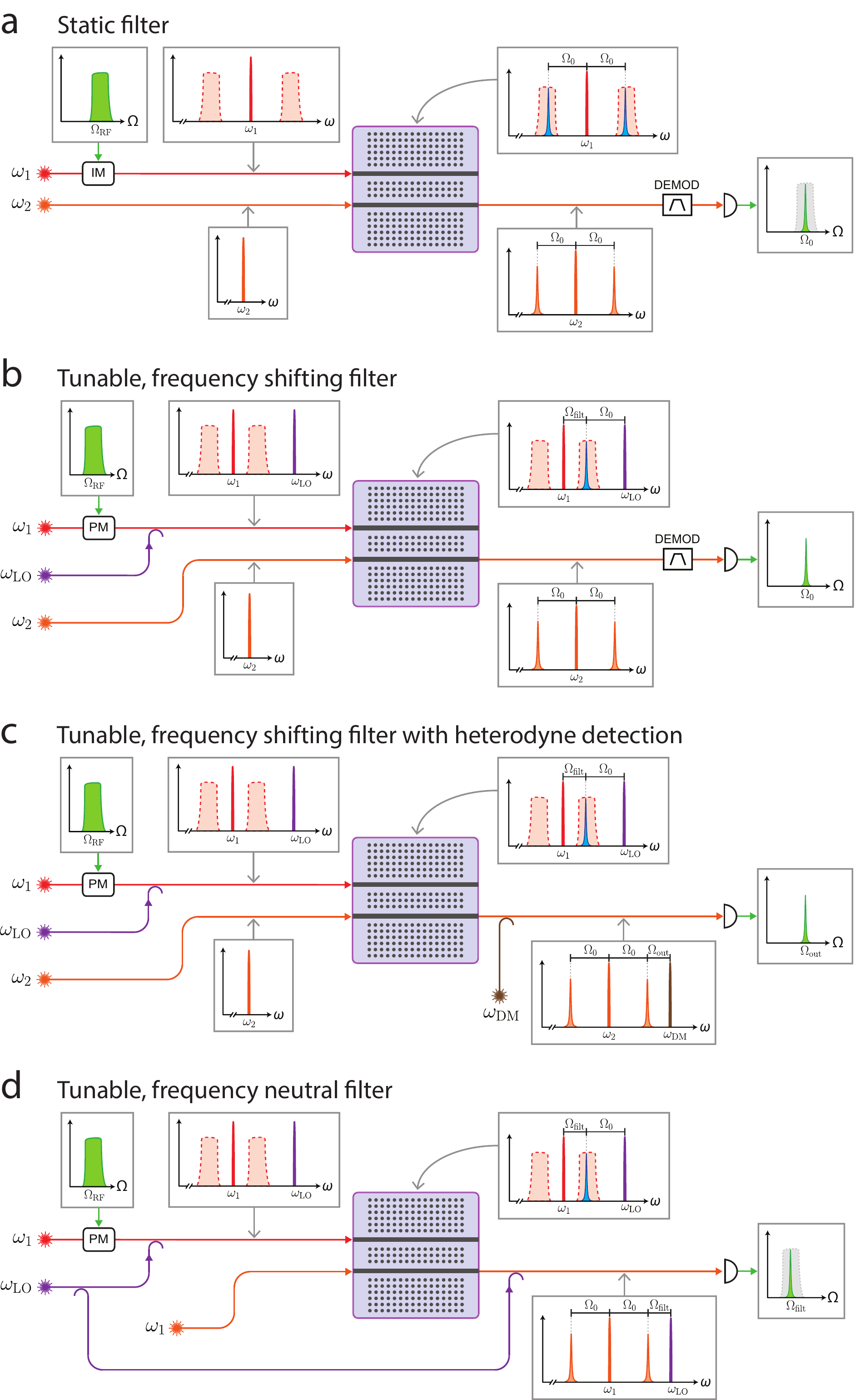}
    \caption{
    \textbf{(a)} Schematic illustration of a PPER-based RF-photonic filter with a set frequency $\Omega_0$, determined by the device geometry.
    \textbf{(b)} By adding a frequency-tunable optical local oscillator (LO) at the link input, the filter pass-band $\Omega_\text{filt}$ can be shifted,  however the output RF signal is set at the Brillouin frequency $\Omega_0$.
    \textbf{(c)} Optical-heterodyne demodulation, using an optical tone at frequency $\omega_\text{DM}$ shifts the output RF signal to $\Omega_\text{out} = \omega_\text{DM} - \omega_2 - \Omega_0$.
    \textbf{(d)} Using the LO to perform heterodyne detection at the link output, and setting the `emit' and `receive' tones to the same frequency ($\omega_2=\omega_1$) yields a frequency-neutral filter, where the output RF signal is at the same frequency as the filter pass-band.
    }
    \label{fig:tunability}
\end{figure*}

When a PPER device is used as a static RF-filter, the Brillouin frequency $\Omega_0$, which is set by the device geometry, determines the pass-band of the filter. Fig. \ref{fig:tunability}(a) shows how the optical carrier in the `emit' waveguide beats with the sidebands produced by the intensity modulator at the filter input, driving the coherent acoustic fields in the device. This induces phase modulation sidebands around the optical tone in the `receive' waveguide, which can be demodulated at the filter output to retrieve the filtered RF signal. The filter frequency can be determined in the device design, by tuning the width of the phononic defect modes, discussed and demonstrated in Section \ref{sec:SI-geom}. Further trimming can be achieved through strain and temperature perturbations which affect the Brillouin frequency.

In order to tune the filter in real time, an alternative modulation scheme can be used at the filter input, demonstrated experimentally in Section \ref{subsec:tune}, and shown in Fig. \ref{fig:tunability}(b). In this scenario, the input signal is phase-modulated onto the optical carrier (optical frequency $\omega_1$) in the `emit' waveguide. Since a phase modulator is used (rather than the intensity modulator in the static filter case), this does not generate a beat-note necessary for the generation of a coherent phonon field. A separate optical tone at optical frequency $\omega_\text{LO}$ is injected into the `emit' waveguide, serving as an optical local oscillator (LO). The LO can generate a beat-note with the `emit' sideband spaced by the Brillouin frequency, and drive an acoustic field. In the example of Fig. \ref{fig:tunability}(b), where the LO is at a higher frequency than the sidebands, the filter pass-band is at frequency $\omega_\text{filt} = \omega_\text{LO} - \Omega_0$, or in the RF domain, $\Omega_\text{filt} = \omega_\text{LO} -\omega_1 - \Omega_0 $ (relative to the optical carrier $\omega_1$).
By tuning the LO frequency, the filter pass band is directly shifted, where the tuning of the LO can be implemented using an optical tunable source or by using an electro-optic modulator. 
The narrow-band phonon field is centered around the Brillouin frequency $\Omega_0$ regardless of the LO frequency, hence the phase modulated sidebands, and the measured RF signal at the filter output (after demodulation) is always at $\Omega_0$, identical to the static filter case. Overall, this yields an RF band-pass filter with a center frequency $\Omega_\text{filt}$, shifted to frequency $\Omega_0$.
Alternatively, the output RF signal can be shifted to a desired frequency (rather than the Brillouin frequency) by employing optical heterodyne demodulation. As shown in \ref{fig:tunability}(c), using an optical tone at frequency $\omega_\text{DM}$ for de-modulation, the resulting RF signal will be at RF frequency $\Omega_\text{out} = \omega_\text{DM}-\omega_2-\Omega_0$, which can be determined by setting $\omega_\text{DM}$ and $\omega_2$.

To achieve a frequency-neutral filter, an alternative demodulation scheme is needed at the RF-photonic link output. In this case, the optical local oscillator is used to perform heterodyne detection to demodulate the phase-modulation sidebands around the `receive' tone exiting the Brillouin-active device. The input of the RF-photonic link is the same as in the tunable frequency shifting filter, such that the pass-band centered around $\Omega_\text{filt}=\omega_\text{LO} - \omega_1 - \Omega_0$ yields phase modulation at the Brillouin frequency $\Omega_0$. Combining the phase-modulated `receive' tone with the LO produces a beat-note between the LO and the phase-modulation sideband, yielding a measurable RF signal. In this example, the RF tone will have frequency $\Omega_\text{out} = \omega_\text{LO} - \omega_2 - \Omega_0$, where $\omega_2$ is the `receive' tone optical carrier. By setting $\omega_2=\omega_1$, i.e. using the same optical frequency in the `emit' and `receive' waveguides, we obtain a frequency neutral filter, where $\Omega_\text{out} = \Omega_\text{filt}$. An illustration of such a frequency-neutral scheme is presented in Fig. \ref{fig:tunability}(d).

\subsection*{\label{subsec:SI-tune-bandwidth}Bandwidth limitations}
\begin{figure*}[hb!]
    \centering
    \includegraphics[scale=0.8]{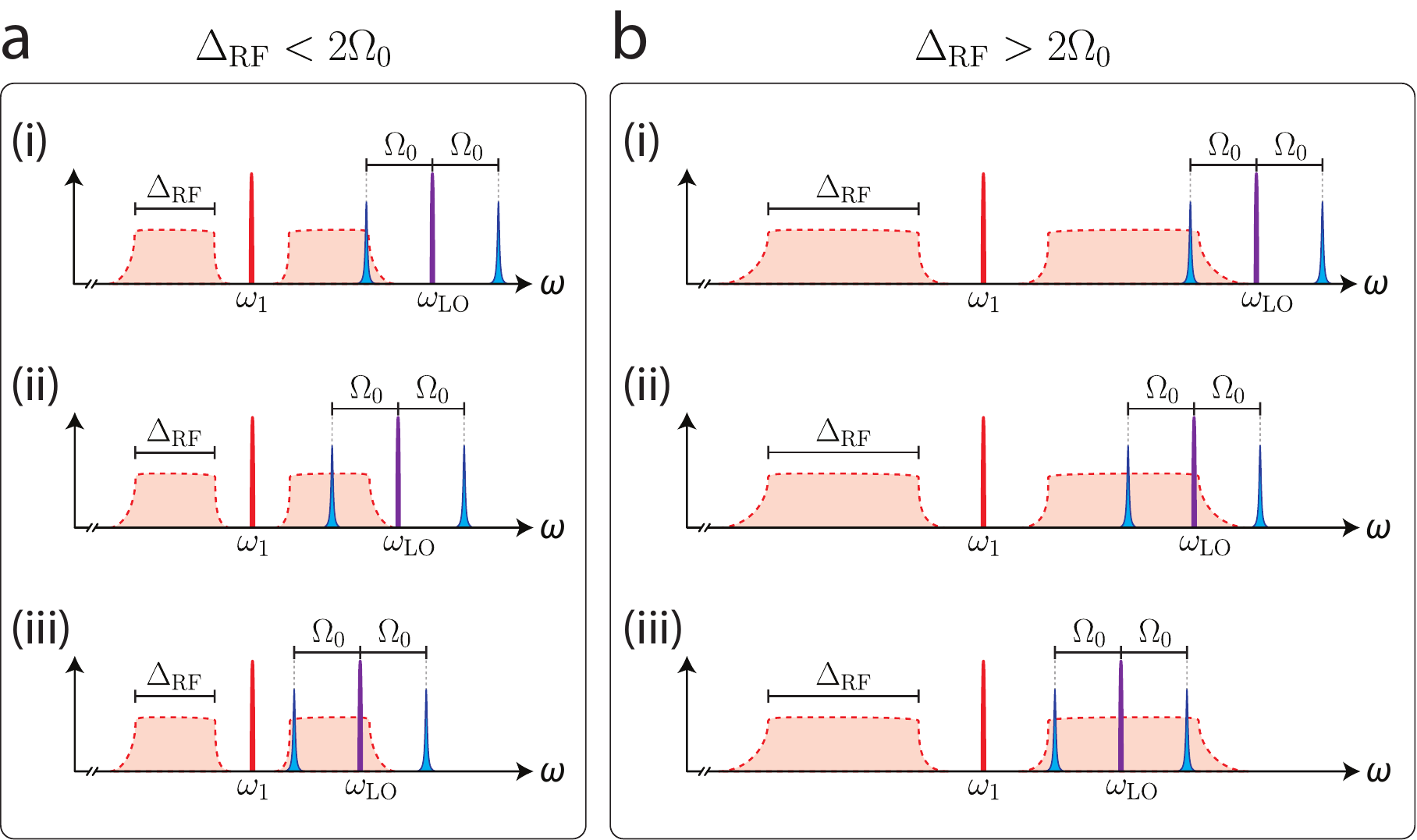}
    \caption{
    Tuning of the optical local oscillator (LO) across the RF-modulated sideband enables filtering of different spectral bands. 
    \textbf{(a)} When the total bandwidth of the RF sideband is smaller than twice the Brillouin frequency ($\Delta_\text{RF}<2\Omega_0$), there is no overlap of the higher LO sideband with the RF-modulated signal, and the filter output corresponds to the transfer function of the acoustic mode with no distortion.
    \textbf{(b)} If the bandwidth limitation is not satisfied ($\Delta_\text{RF} > 2\Omega_0$), as we tune the pass-band of the filter, both sidebands of the LO can overlap with the RF signal (panel ({\romannumeral 3})), yielding distortion at the filter output, equivalent to image-frequency interference.
    }
    \label{fig:bandwidth}
\end{figure*}

The forward-Brillouin scattering used in the PPER-based filtering scheme is inherently double-sided, such that the Stokes (red-shift) and anti-Stokes (blue-shift) scattering processes are coupled \cite{kang2009tightly,kittlaus2016large,gertler2019shaping}. 
In the static-filter case (Fig.\ref{fig:tunability}(a)) this results in the carrier beating with both sidebands in the `emit' waveguide, and in fact yields higher transduction efficiency.
However, in the frequency-tuning scheme presented here, this results in the fact that the optical local oscillator (LO) can drive the acoustic field by beating with light spaced by the Brillouin frequency at both higher $\omega_\text{LO} + \Omega_0$ and lower $\omega_\text{LO} - \Omega_0$ frequencies. This is illustrated in Fig. \ref{fig:bandwidth}, showing the two sidebands (blue) where forward-Brillouin scattering can occur around the LO (purple). 
The dual-sideband nature of the process is equivalent to an image-frequency in heterodyne detection, where unwanted frequency components interfere with the desired signal \cite{mirabbasi2000classical}.  

As long as the input-signal RF bandwidth $\Delta_\text{RF}$ is smaller than twice the Brillouin frequency $\Delta_\text{RF}<2\Omega_0$, there will not be distortion in the filter operation. As shown in Figs. \ref{fig:bandwidth}(a), when tuning the LO, only the lower `LO sideband' overlaps with the modulated information. 
It is important to note that this RF bandwidth corresponds to the spectral contents, i.e. $\Delta_\text{RF}=\max(\Omega_\text{RF})-\min(\Omega_\text{RF})$, and does not constrain the highest frequency allowed at the input.
In contrast, when this bandwidth limitation is not satisfied, such that $\Delta_\text{RF} > 2\Omega_0$, the measured output signal will not always be proportional to the filtered input RF signal. This is illustrated in Fig. \ref{fig:bandwidth}(b) panel ({\romannumeral 3}), showing that as the LO approaches the center of the RF sideband it drives the acoustic field through two scattering process whose contributions will be summed.
This distortion can be avoided by using an image-rejection filter, common in many homodyne Rf receiver schemes \cite{mirabbasi2000classical}, ensuring single-sideband filtering. Alternatively, this bandwidth limitation can be lifted by using inter-band Brillouin-scattering which is inherently a single-sideband process  \cite{kittlaus2017_Intermodal,kittlaus2018non}.

We note that throughout this analysis we are assuming the small-signal regime, such that the RF-modulation sidebands are much smaller than the optical carrier and the LO. This enables us to neglect the beat-note generated between different spectral components within and between the RF sidebands, as their contribution will be much smaller than that of the beat-note generated by the strong LO.

\clearpage
\section{\label{sec:SI-multi_pole}Multi-pole frequency response}
\subsection*{\label{subsec:SI-multi_pole-2_pole}Second-order filters}
\begin{figure*}[htb!]
    \centering
    \includegraphics[scale=0.77]{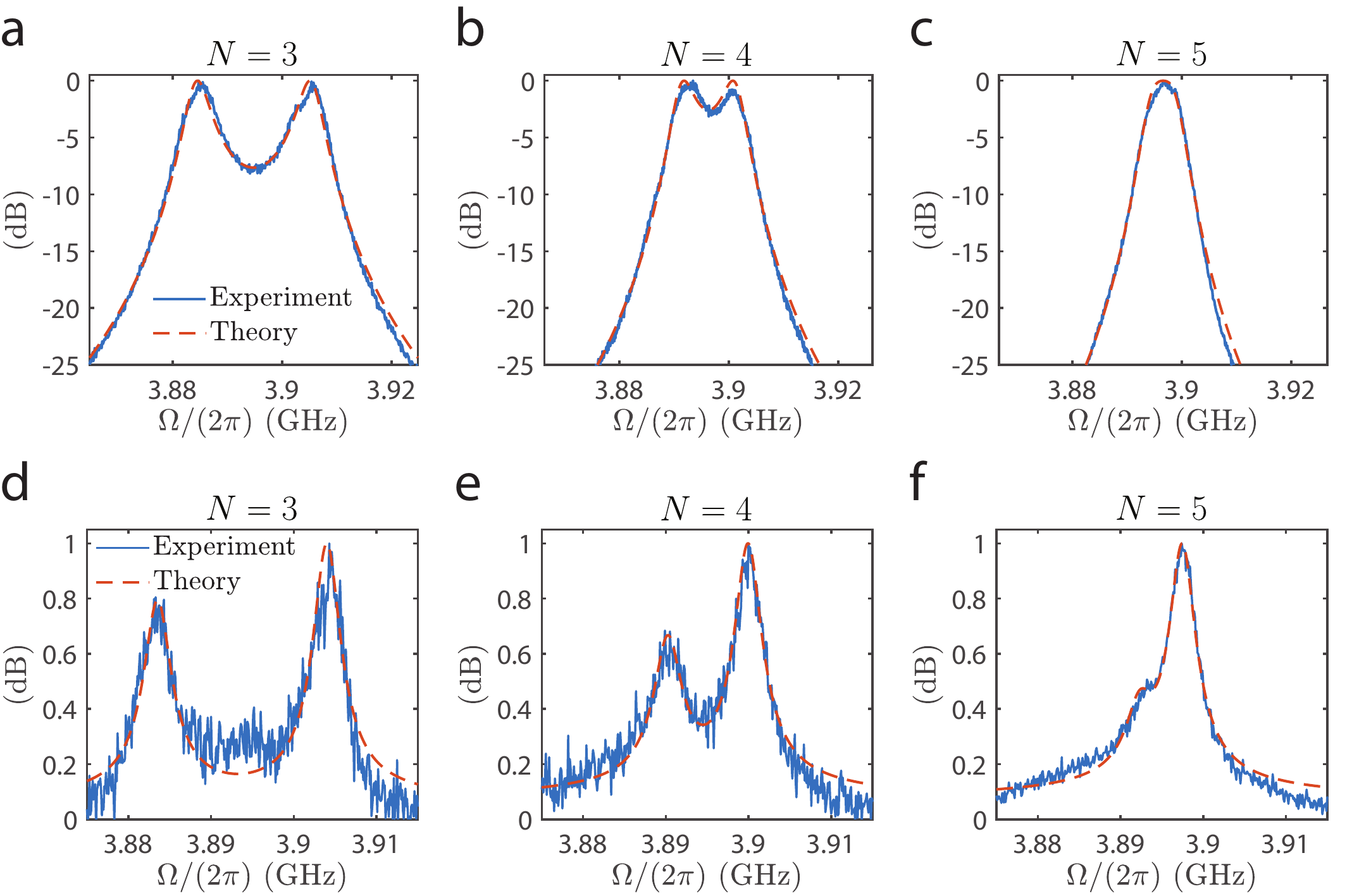}
    \caption{
    \textbf{(a-c)} Normalized measured frequency response of PPER filter responses with $N=3,4,5$ rows of holes between the two acoustic defect regions, respectively.
    \textbf{(d-f)} Normalized measurements of spontaneous-Brillouin scattering in the same three devices, revealing the thermal occupation of the two super-modes. The phononic crystal design used in these devices has a pitch of $a = 600$ nm and hole diameter of $d=462$ nm, with a defect width $W=3.3\ \mu$m.
    }
    \label{fig:2pole}
\end{figure*}

The frequency response of the two-pole PPER filter is a result of the coherent interaction of the two acoustic modes taking part in the signal transduction \cite{gertler2019shaping}, with a magnitude given by
\begin{equation}
    \left|\chi^\text{(2 pole)}(\Omega)\right|^2 \propto \frac{\mu^2}{ \left[\left(\Gamma/2\right)^2+\left(\Omega-\Omega_0+\mu\right)^2\right] \left[\left(\Gamma/2\right)^2+\left(\Omega-\Omega_0-\mu\right)^2\right]},
    \label{eq:2pole_mag}
\end{equation}
where $\mu$ is the coupling rate between the two acoustic modes, $\Gamma$ the acoustic dissipation rate, and $\Omega_0$ the resonant frequency of each of the two separate acoustic modes, which we assume are equal for both acoustic modes. 
Figs. \ref{fig:2pole}(a-c) show the measured frequency response of three devices with a resonant frequency of 3.895 GHz, each with a different coupling rate. This was achieved by fabricating devices with $N=3,4,5$ rows of holes between the two acoustic defect regions, yielding fitted coupling rates of $\mu/(2\pi) = 10.5,5,2$ MHz respectively, and acoustic dissipation rates of $\Gamma/(2\pi) = 4.6,4.4,5.1$ MHz. The two peaks in the frequency response are separated by $\delta\Omega=2\mu[{1-(\Gamma/(2\mu))^2}]^{1/2}$, which can be described as the frequency splitting resulting from the coupling of two degenerate acoustic modes, or equivalently, the frequencies of the two eigen-modes of the coupled system. When the coupling is smaller than half the dissipation rate ($\mu < \Gamma/2$), this yields a single-peaked line shape. 
The super-modes are linear combinations of the two spatial acoustic modes, given by \cite{gertler2019shaping}
\begin{equation}
    b_{+} = \left(b_\text{A} + b_\text{B}\right) /\sqrt{2}, \qquad
    b_{-} = \left(b_\text{A} - b_\text{B}\right) /\sqrt{2},
    \label{eq:2pole_evect}
\end{equation}
where we have denoted the two spatially separated acoustic modes as $b_\text{A}$ and $b_\text{B}$.
These frequency responses exhibit a sharp frequency roll-off, with a similar response outside of the pass-band, regardless of the acoustic coupling rate.

Spontaneous scattering also occurs in the PPER devices, as result of the thermal occupation of the acoustic modes at room temperature \cite{kharel2016_Hamiltonian}. The power-spectrum of these fluctuations can be described by summing the contribution of the acoustic super-modes of the device \cite{gertler2019shaping}, given by
\begin{equation}
    \left|\chi_N(\Omega)\right|^2 = \frac{1/2}{\left(\Gamma/2\right)^2+\left(\Omega-\Omega_0+\mu\right)^2} + \frac{1/2}{\left(\Gamma/2\right)^2+\left(\Omega-\Omega_0-\mu\right)^2},
    \label{eq:2pole_N}
\end{equation}
where we have assumed two identical acoustic modes.

Figs. \ref{fig:2pole}(d-f) show measured spontaneous-Brillouin scattering of the same three devices, demonstrating the frequency response for different coupling rates. The fitted coupling rates from the spontaneous measurements are $\mu/(2\pi) = 10.1,4.6,2$ MHz, in good agreement with the measurements from Figs. \ref{fig:2pole}(a-c).
The asymmetry in the magnitude of the two peaks, deviating from Eq. (\ref{eq:2pole_N}), is a result of asymmetry between the two coupled acoustic modes, discussed further in Section \ref{sec:SI-asym}.
\clearpage
\subsection*{\label{subsec:SI-multi_pole-3_pole}Third-order filters}
The PPER-based filtering scheme can be extended to higher-order filters by coupling more acoustic modes \cite{gertler2019shaping}. For example, by coupling three spatially separated modes $b_\text{A}$, $b_\text{B}$ and $b_\text{C}$, the frequency response of the PPER operation will be
\begin{equation}
    \left|\chi^\text{(3 pole)}(\Omega)\right|^2 \propto \frac{\mu^4}{ \left[\left(\Gamma/2\right)^2+\left(\Omega-\Omega_0+\sqrt{2}\mu\right)^2\right]\left[\left(\Gamma/2\right)^2+\left(\Omega-\Omega_0\right)^2\right]\left[\left(\Gamma/2\right)^2+\left(\Omega-\Omega_0-\sqrt{2}\mu\right)^2\right]},
    \label{eq:3pole_mag}
\end{equation}
where the three super-modes of the coupled system are given by
\begin{equation}
    b_{-} = \left(b_\text{A} - \sqrt{2} \ b_\text{B} + b_\text{C}\right) /2, \qquad
    b_{0} = \left(b_\text{A} - b_\text{C}\right) /\sqrt{2}, \qquad
    b_{+} = \left(b_\text{A} + \sqrt{2} \ b_\text{B} + b_\text{C}\right) /2.
    \label{eq:3pole_evect}
\end{equation}

\begin{figure*}[hb!]
    \centering
    \includegraphics[scale=0.77]{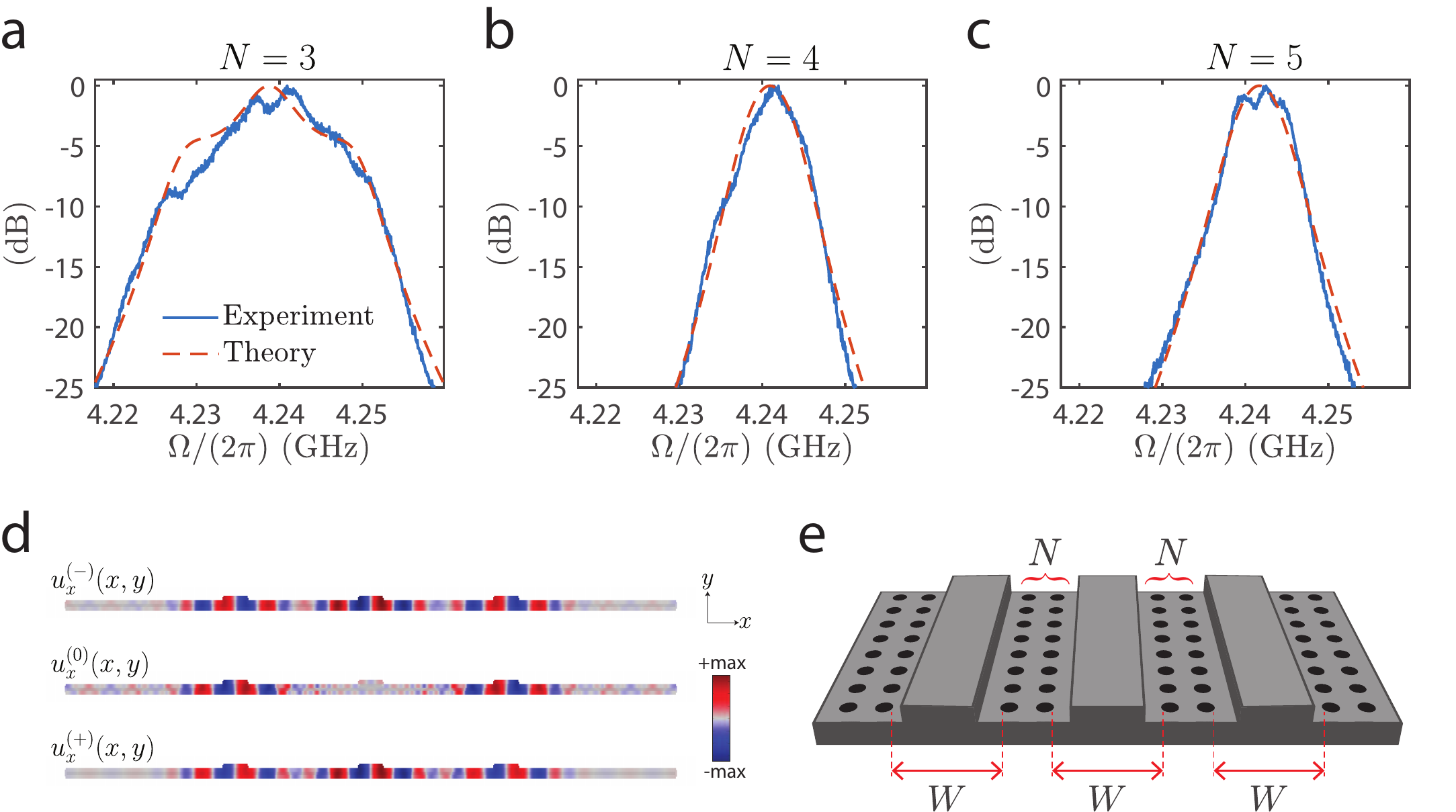}
    \caption{
    \textbf{(a-c)} Normalized measured frequency response of three-pole PPER filter responses with $N=3,4,5$ rows of holes between the acoustic defect regions, respectively.    
    The phononic crystal design used in these devices had a pitch of $a=600$ nm, hole diameter of $d=462$ nm, and defect widths of $W=3\ \mu$m.
    \textbf{(d)} FEM simulation of the $x$-component of the displacement $u_x$ of the three acoustic super-modes taking part in the PPER operation.
    \textbf{(e)} An illustration of a three-pole device, showing the three defect regions of width $W$, with nearest-neighbor coupling through $N$ rows of holes.
    }
    \label{fig:3pole}
\end{figure*}

A three-pole PPER device is schematically illustrated in Fig. \ref{fig:3pole}(e), where three phonon defect modes of width $W$ are coupled through $N$ rows of holes.
Simulation of the acoustic super-modes supported by the structure are presented in Fig. \ref{fig:3pole}(d), consistent with Eq. (\ref{eq:3pole_evect}).
Measured frequency responses of fabricated three-pole PPER devices are presented in Figs. \ref{fig:3pole}(a-c), corresponding to $N=3,4,5$ rows of holes between each of the acoustic defect regions, respectively. The fitted frequency responses yield coupling rates of $\mu/(2\pi) = 7.5,2.1,1.8$ MHz respectively, and acoustic dissipation rates of $\Gamma/(2\pi) = 7,7.7,9.3$ MHz.
The non-ideal measured line shapes can be a consequence of the larger device structure, resulting in a wider suspended silicon region and the need to define more holes in the phononic crystal. This can lead to a larger degree of non-uniformity in the fabrication process, and a deviation from the theoretical functions.
The devices presented in this work were all fabricated using a two-step electron-beam lithography process (see Section \ref{sec:SI-fab} for more details) which can suffer from drift, however implementation of these designs using photo-lithography may yield higher uniformity and more ideal line shapes.

\clearpage
\section{\label{sec:SI-asym}Effects of acoustic mode asymmetry}
The line shape of PPER-based fitlers depends on the acoustic resonant frequencies and coupling rates, which are determined by their geometry. The two acoustic modes are designed to be identical, however fabrication imperfections and material non-uniformity can result in an asymmetry between the modes. This can lead to variations in the resonant frequency and acousto-optic coupling rate. For example, a perturbation of 10 nanometers can result in a deviation of resonant frequency on the order of 10 MHz \cite{kittlaus2016large}.

To take asymmetry into account, we describe the acoustic modes using two different frequencies for each of the spatial acoustic modes. Following the description in Ref. \cite{gertler2019shaping}, we can write the acoustic part of the system Hamiltonian with two coupled acoustic modes with resonant frequencies $\Omega_0^\text{(A)}$ and $\Omega_0^\text{(B)}$, and a coupling rate $\mu$
\begin{equation}
    H_{\text{ac}} = \hbar\int{dz\ \begin{pmatrix} {b^\text{(A)}}^{\dagger} & {b^\text{(B)}}^{\dagger} \end{pmatrix} 
    \begin{pmatrix} {\Omega}_0^\text{(A)} & \mu & \\
    \mu^* & {\Omega}_0^\text{(B)} \end{pmatrix} 
    \begin{pmatrix} b^\text{(A)} \\ b^\text{(B)} \end{pmatrix}}
    \label{eq:2_mode_H}.
\end{equation}

The matrix in Eq. (\ref{eq:2_mode_H}) is Hermitian, and we can diagonlize it using a unitary matrix $V$ such that
\begin{equation}
    \begin{pmatrix} {\Omega}_+ & 0 & \\
    0 & {\Omega}_- \end{pmatrix}=
    V^\dagger  
    \begin{pmatrix} {\Omega}_0^\text{(A)} & \mu & \\
    \mu^* & {\Omega}_0^\text{(B)} \end{pmatrix} 
    V,
\end{equation}
where $\Omega_\pm$ are the eigen-frequencies of the coupled system, and the eigen-modes are given by
\begin{equation}
    \begin{pmatrix} b_+ \\ b_- \end{pmatrix} = V^\dagger \begin{pmatrix} b^\text{(A)} \\ b^\text{(B)} \end{pmatrix}.
    \label{eq:eig_freq_V}
\end{equation}

The eigen-frequencies can be directly calculated
\begin{equation}
    {\Omega}_\pm = {\Omega}_0 \pm \sqrt{\left(\frac{\Delta\Omega}{2}\right)^2 + \left|\mu\right|^2},
    \label{eq:eig_freq}
\end{equation}
where we define the average frequency ${\Omega}_0 = (\Omega_0^\text{(A)}+\Omega_0^\text{(B)})/2$ and the frequency difference $\Delta\Omega = {\Omega}_0^\text{(B)}-{\Omega}_0^\text{(A)}$. The elements of the unitary matrix $V$ are given by
\begin{equation}
    V_{\pm}^\text{(A)} = \frac{1}{\mathcal{N}_\pm} \ \left(-\frac{\Delta\Omega}{2} \pm \sqrt{\left(\frac{\Delta\Omega}{2}\right)^2 + \left|\mu\right|^2}\right),\qquad
    V_{\pm}^\text{(B)} = \frac{1}{\mathcal{N}_\pm} \ \mu,
\label{eq:eig_modes}
\end{equation}
with the normalization factor
\begin{equation}
    \mathcal{N}_\pm = \left|\mu\right|\sqrt{ 1+\left| \frac{\Delta\Omega}{2 |\mu|} \mp \sqrt{1+\left( \frac{\Delta\Omega}{2 |\mu|} \right)^2} \ \right|^2 }.
    \label{eq:eig_modes_norm}
\end{equation}

\begin{figure*}[hb!]
    \centering
    \includegraphics[scale=0.77]{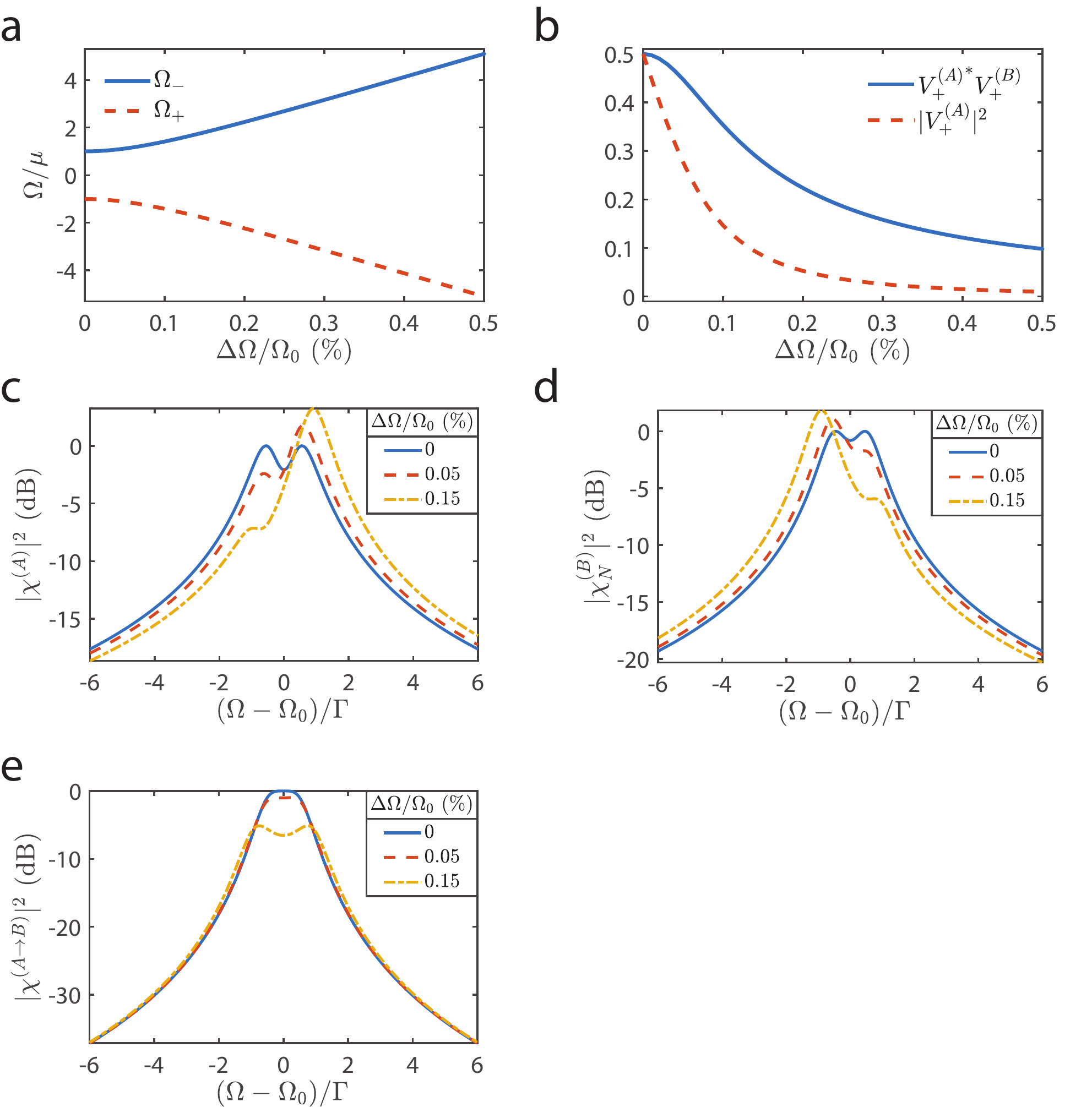}
    \caption{
    \textbf{(a)} Calculated eigen-frequencies of a PPER device, as a function of the asymmetry between the two acoustic modes, $\Delta\Omega = {\Omega}_0^\text{(B)}-{\Omega}_0^\text{(A)}$.
    \textbf{(b)} The coefficients determining the frequency response (Eq. (\ref{eq:eig_modes})) as a function of asymmetry between the acoustic modes.
    \textbf{(c)} The frequency response of the phase-modulation experienced by the field propagating in the `emit' waveguide, for different amounts of asymmetry.
    \textbf{(d)} The power spectrum of spontaneous-Brillouin scattering in the `receive' waveguide, for different amounts of asymmetry.
    \textbf{(e)} The filter shape of a PPER operation, for different amounts of asymmetry.
    The calculations assume $\Omega_0/\Gamma = 1000$ and $\mu=\Gamma/2$.
    }
    \label{fig:asym_theory}
\end{figure*}

Since the diagonalizing matrix $V$ is unitary, the matrix elements follow the relations
\begin{equation}
\begin{split}
\left|V_{+}^\text{(A)}\right|^2 + \left|V_{-}^\text{(A)}\right|^2 = 1,\qquad
\left|V_{+}^\text{(B)}\right|^2 + \left|V_{-}^\text{(B)}\right|^2 = 1,\qquad
V_{\pm}^\text{(B)} {V_{\pm}^\text{(A)}}^* + V_{\mp}^\text{(B)} {V_{\mp}^\text{(A)}}^* = 0,
\end{split}
\label{eq:orthnormality}
\end{equation}
which is equivalent to the ortho-normality of the eigen-basis. Figs. \ref{fig:asym_theory}(a) and (b) show an example of the dependence of the eigen-frequencies and the coefficients $V^{(\ell)}_{\pm}$ on the asymmetry between the two acoustic modes.
We can see that in the case of two identical modes, the frequency splitting is exactly twice the coupling rate, and the coefficients are equal in magnitude $|V^{(\ell)}_{\pm}|=1/\sqrt{2}$.

The equations of motion describing the optical and acoustic modes in the two waveguides of the PPER, denoted A and B, are given by \cite{gertler2019shaping}
\begin{equation}
\begin{split}
b_{\pm} &= \left(\frac{1}{\Omega - \Omega_{\pm} + i\frac{\Gamma}{2}} \right) \sum_n\left({g_{\pm}^\text{(A)}}^* a_{n}^\text{(A)} {a_{n-1}^\text{(A)}}^{\dagger}+{g_{\pm}^\text{(B)}}^* a_{n}^\text{(B)} {a_{n-1}^\text{(B)}}^{\dagger} \right),\\
\frac{\partial {a}_{n}^\text{(A)}}{\partial z} &= -\frac{i}{v} \Big( g_{+}^\text{(A)} {a}_{n-1}^\text{(A)} b_{+} + {g_{+}^\text{(A)}}^* {a}_{n+1}^\text{(A)} b^{\dagger}_{+} + g_{-}^\text{(A)} {a}_{n-1}^\text{(A)} b_{-} + {g_{-}^\text{(A)}}^* {a}_{n+1}^\text{(A)} b^{\dagger}_{-} \Big),\\
\frac{\partial {a}_{n}^\text{(B)}}{\partial z} &= -\frac{i}{v} \Big( g_{+}^\text{(B)} {a}_{n-1}^\text{(B)} b_{+} + {g_{+}^\text{(B)}}^* {a}_{n+1}^\text{(B)} b^{\dagger}_{+} + g_{-}^\text{(B)} {a}_{n-1}^\text{(B)} b_{-} + {g_{-}^\text{(B)}}^* {a}_{n+1}^\text{(B)} b^{\dagger}_{-} \Big),
\end{split}
\label{eq:FSBS_two_membranes}
\end{equation}
where $v$ is the optical group velocity of the guided optical modes, and we have absorbed the coefficients $V$ into the acousto-optic coupling rates $g$, such that $g^{(\ell)}_{\pm} = g^{(\ell)} {V^{(\ell)}_{\pm}}$.

Following the derivation in Ref. \cite{gertler2019microwave}, the phonon fields in the PPER structure are given by
\begin{equation}
    b_\pm \propto -i\chi_\pm {g_\pm^\text{(A)}}^* \left|{a}^\text{(A)}_{0}(0)\right|^2,
    \label{eq:phonon_field_two_tones}
\end{equation}
where we have defined the frequency responses $\chi_\pm = [i(\Omega_\pm-\Omega)+\Gamma/2]^{-1}$, and the proportionality constant takes into account the details of the modulation scheme at the input of waveguide A. Plugging back into the equations of motion of the optical fields yields
\begin{equation}
\begin{split}
\frac{\partial {a}_{n}^\text{(A)}}{\partial z} &\propto -\frac{1}{v} \left|{a}^\text{(A)}_{0}(0)\right|^2 \Big[ {a}_{n-1}^\text{(A)} e^{i\Lambda} \left(\chi_+ \left|g_+^\text{(A)}\right|^2 + \chi_- \left|g_-^\text{(A)}\right|^2 \right) - {a}_{n+1}^\text{(A)} e^{-i\Lambda} \left(\chi_+^* \left|g_+^\text{(A)}\right|^2 +\chi_-^* \left|g_-^\text{(A)}\right|^2\right)\Big],\\
\frac{\partial {a}_{n}^\text{(B)}}{\partial z} &\propto -\frac{1}{v} \left|{a}^\text{(A)}_{0}(0)\right|^2 \Big[ {a}_{n-1}^\text{(B)} e^{i\Lambda} \left(\chi_+ g_+^\text{(B)} {g_+^\text{(A)}}^* + \chi_- g_-^\text{(B)} {g_-^\text{(A)}}^* \right) - {a}_{n+1}^\text{(B)} e^{-i\Lambda} \left(\chi_+^* {g_+^\text{(B)}}^* g_+^\text{(A)} +\chi_-^* {g_-^\text{(B)}}^* g_-^\text{(A)}\right)\Big].
\end{split}
\label{eq:FSBS_two_membranes_no_b}
\end{equation}

We factor out the single waveguide coupling rate $g^{(l)}_{\pm} = g^{(l)} {V^{(l)}_{\pm}}$, leaving us with
\begin{equation}
\begin{split}
\frac{\partial {a}_{n}^\text{(A)}}{\partial z} &\propto -\frac{1}{v} \left|{a}^\text{(A)}_{0}(0)\right|^2\left|g^\text{(A)}\right|^2 \left| \chi^\text{(A)} \right| \left( {a}_{n-1}^\text{(A)} e^{i\phi^\text{(A)}} - {a}_{n+1}^\text{(A)} e^{-i\phi^\text{(A)}}\right),\\
\frac{\partial {a}_{n}^\text{(B)}}{\partial z} &\propto -\frac{1}{v} \left|{a}^\text{(A)}_{0}(0)\right|^2\left|g^\text{(B)} {g^\text{(A)}}^*\right|^2 \left| \chi^ {\text{(A}\rightarrow \text{B)}}\right| \left( {a}_{n-1}^\text{(B)} e^{i\phi^{\text{(A}\rightarrow \text{B)}}}  - {a}_{n+1}^\text{(B)} e^{-i\phi^{\text{(A}\rightarrow \text{B)}}} \right),
\end{split}
\label{eq:FSBS_two_membranes_no_b_2}
\end{equation}
where we have also defined the frequency response in each waveguide
\begin{equation}
\chi^\text{(A)} =\Bigg[\chi_+ \left|V_+^\text{(A)}\right|^2 + \chi_- \left|V_-^\text{(A)}\right|^2 \Bigg],\qquad
\chi^{\text{(A}\rightarrow \text{B)}} =\Bigg[\chi_+ V_+^\text{(B)} {V_+^\text{(A)}}^* + \chi_- V_-^\text{(B)} {V_-^\text{(A)}}^* \Bigg],
\label{eq:freq_res_asymm}
\end{equation}
and denoted the phase responses $\phi^\text{(A)} = \arg\left(\chi^\text{(A)}\right)$, $\phi^{\text{(A}\rightarrow \text{B)}} = \arg\left(\chi^{\text{(A}\rightarrow \text{B)}}\right)$. Using the relations from Eq. (\ref{eq:orthnormality}) we can rewrite these frequency responses
\begin{equation}
\chi^\text{(A)} =\Bigg[\chi_+ \left|V_+^\text{(A)}\right|^2 + \chi_- \left( 1 - \left|V_+^\text{(A)}\right|^2 \right) \Bigg],\qquad
\chi^{\text{(A}\rightarrow \text{B)}} = V_+^\text{(B)} {V_+^\text{(A)}}^* \Big[\chi_+  - \chi_- \Big].
\label{eq:freq_res_asymm_2}
\end{equation}

The frequency response $\chi^{\text{(A}\rightarrow \text{B)}}$ describes the filter line shape obtained in the PPER operation (transducing information from waveguide A to B), whereas $\chi^\text{(A)}$ describes the frequency response of the phase modulation experienced by the optical field propagating through the `emit' waveguide \cite{gertler2019shaping}.
Figs. \ref{fig:asym_theory}(c) and (e) present these frequency responses for different amounts of asymmetry in the two acoustic modes. The filter response $\chi^{\text{(A}\rightarrow \text{B)}}$ stays symmetric around the pass-band center, even in the case of non-identical acoustic modes. In contrast, the function $\chi^\text{(A)}$ shows increasing asymmetry for larger values of $\Delta\Omega$.

The frequency response of spontaneous scattering can be calculated by an incoherent sum of the two super modes \cite{gertler2019shaping}. The resulting power-spectrum of these thermal fluctuations in waveguide B (which can determine the noise-floor of a PPER-based photonic filter) are given by
\begin{equation}
\left|\chi^\text{(B)}_N\right|^2 = \Bigg[\left|\chi_+ \right|^2 \left|V_+^\text{(B)}\right|^2 + \left|\chi_- \right|^2 \left( 1 - \left|V_+^\text{(B)}\right|^2 \right) \Bigg],
\label{eq:freq_res_noise}
\end{equation}
showing asymmetry in the frequency response, as demonstrated in Fig. \ref{fig:asym_theory}(d). Both of the frequency responses $\chi^\text{(A)}$ and $\chi^\text{(B)}_N$ will be mirrored across the center frequency when using the opposite waveguide, i.e. looking at the responses $\chi^\text{(B)}$ and $\chi^\text{(A)}_N$. In contrast, the PPER filter frequency response $\chi^{\text{(A}\rightarrow \text{B)}}$, is symmetric, such that exchanging the roles of waveguides A and B in the filtering operation will not alter the line shape.

\vspace{2mm}

\begin{figure*}[hb!]
    \centering
    \includegraphics[scale=0.77]{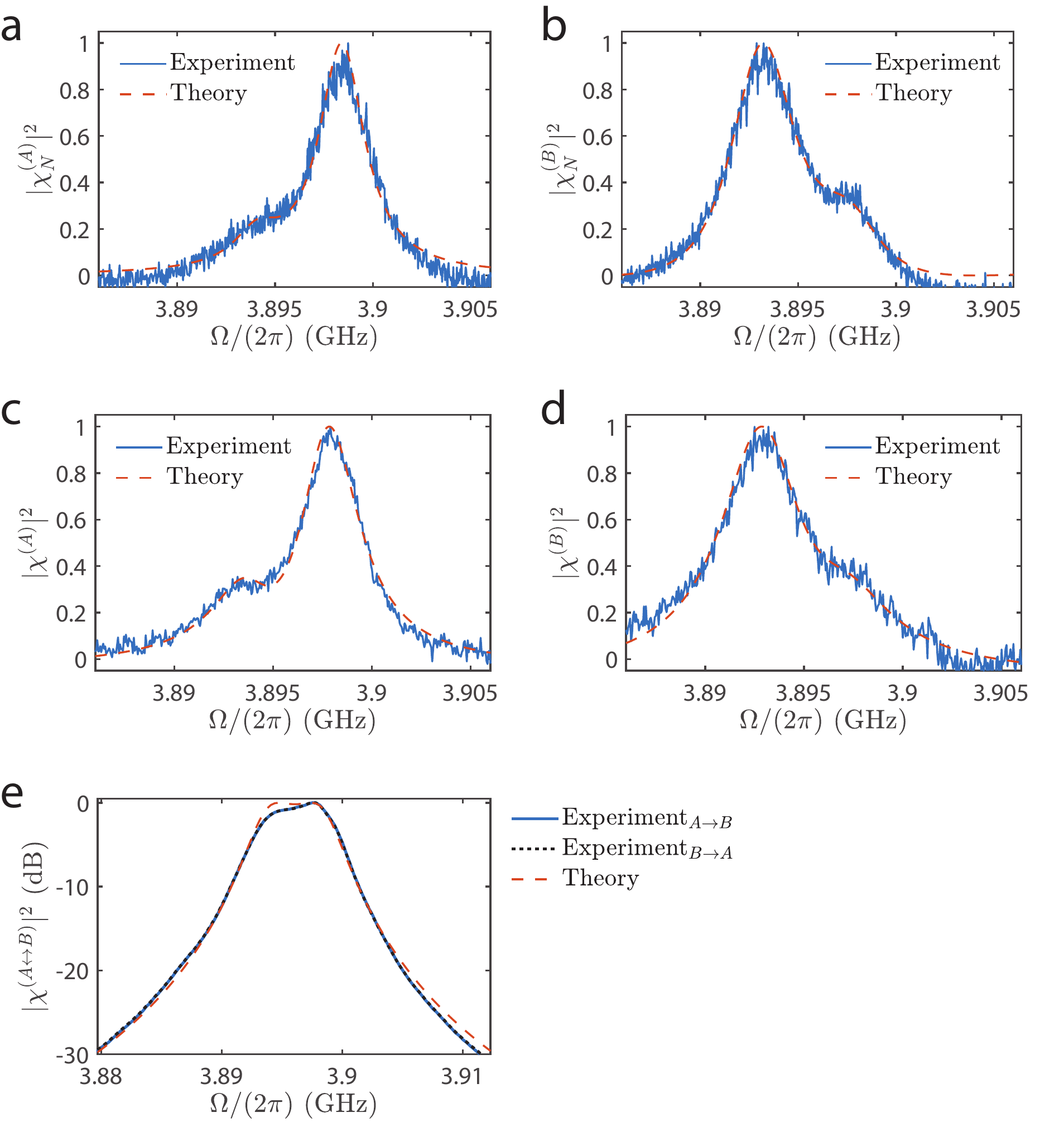}
    \caption{
    \textbf{(a)} Normalized power spectrum of spontaneous Brillouin scattering in the `emit' waveguide.
    \textbf{(b)} Normalized power spectrum of spontaneous Brillouin scattering in the `receive' waveguide, showing a mirrored line shape.
    \textbf{(c)} Normalized frequency response of the phase modulation experienced by an optical field propagating in the `emit' waveguide.
    \textbf{(d)} Normalized frequency response of the phase modulation experienced by  an optical field propagating in the `receive' waveguide, showing a mirrored line shape.
    \textbf{(e)} The two-pole filter shape of a PPER operation, for both possible choices of `emit' and `receive' waveguides.
    }
    \label{fig:asym_data}
\end{figure*}

Fig. \ref{fig:asym_data} presents measured frequency responses of spontaneous scattering (panels (a), (b)), and forward Brillouin-induced phase modulation (panels (c), (d)). 
As can be seen from the data, the asymmetric line shapes are mirrored through the center of the trace when switching between waveguides A and B, consistent with Eqs. (\ref{eq:freq_res_asymm_2}) and (\ref{eq:freq_res_noise}). The measurements are consistent with an asymmetry of $\Delta\Omega=2.65 \pm 0.5$ MHz, corresponding to a deviation of $0.03\%$ per waveguide from the average frequency, which can be a result of variations on the order of $\sim2$ nanometers in the geometry of each of the two acoustic guiding structures.
In contrast, Fig. \ref{fig:asym_data}(e) shows that the PPER filter frequency response does not change when switching the roles of `emit' and `receive' between waveguides A and B, consistent with  Eq. (\ref{eq:freq_res_asymm_2}).

\clearpage
\section{\label{sec:SI-xtalk}Effects of optical cross-talk}
In our analysis, we have assumed that there is no optical cross-talk between the two waveguides. This is a good approximation, as the optical guiding regions of the waveguides are separated by $\sim4\ \mu$m, larger than the optical wavelength scale. However, small residual coupling, or optical coupling occurring in other parts of the chip can have an effect on the filter line shape obtained in a PPER operation.
Using Eq. (\ref{eq:FSBS_two_membranes}), and assuming optical cross-talk $\varepsilon$ from waveguide A to B, the phonon field can be described by
\begin{equation}
    b_\pm \propto -i \chi_\pm \left( {g_\pm^\text{(A)}}^* + \varepsilon {g_\pm^\text{(B)}}^*\right) \left|{a}^\text{(A)}_{0}(0)\right|^2,
    \label{eq:phonon_field_two_tones_w_cross_talk}
\end{equation}
where we have assumed $\left|\varepsilon\right| \ll 1$ such that we can neglect the energy lost in waveguide A, and we have denoted the frequency response $\chi_\pm = [i(\Omega_\pm-\Omega)+\Gamma/2]^{-1}$. 
Plugging back into Eq. (\ref{eq:FSBS_two_membranes}) gives us the equation of motion for the optical field in waveguide B
\begin{equation}
\begin{split}
\frac{\partial {a}_{n}^\text{(B)}}{\partial z} \propto -\frac{1}{v} \left|{a}^\text{(A)}_{0}(0)\right|^2 \Bigg( {a}_{n-1}^\text{(B)} & \left[ \chi_+\left( g_+^\text{(B)} {g_+^\text{(A)}}^* + \varepsilon \left|g_+^\text{(B)}\right|^2\right)+ \chi_- \left(g_-^\text{(B)} {g_-^\text{(A)}}^* + \varepsilon \left|g_-^\text{(B)}\right|^2\right) \right] \\
& - {a}_{n+1}^\text{(B)} \left[\chi_+^* \left({g_+^\text{(B)}}^* g_+^\text{(A)} + \varepsilon^* \left|g_+^\text{(B)}\right|^2 \right)+\chi_-^* \left({g_-^\text{(B)}}^* g_-^\text{(A)} \varepsilon^* \left|g_-^\text{(B)}\right|^2 \right)\right]\Bigg).
\end{split}
\end{equation}

We again factor out the rate $g$, yielding
\begin{equation}
\begin{split}
\frac{\partial {a}_{n}^\text{(B)}}{\partial z} \propto -\frac{1}{v} \left|g\right|^2 \left|{a}^\text{(A)}_{0}(0)\right|^2 \left|\widetilde{\chi}^{\text{(A}\rightarrow \text{B)}} \right| \left( {a}_{n-1}^\text{(B)}  e^{i \widetilde{\phi}^\text{(B)}} - {a}_{n+1}^\text{(B)} e^{-i \widetilde{\phi}^\text{(B)}}\right),
\end{split}
\end{equation}
where the frequency response is given by
\begin{equation}
\widetilde{\chi}^{\text{(A}\rightarrow \text{B)}} =\Bigg[ \left( V_+^\text{(B)} {V_+^\text{(A)}}^* + \varepsilon \left|V_+^\text{(B)}\right|^2\right) \chi_+ + \left(V_-^\text{(B)} {V_-^\text{(A)}}^* + \varepsilon \left|V_-^\text{(B)}\right|^2\right) \chi_- \Bigg],
\end{equation}
and we have defined \(\widetilde{\phi}^{\text{(A}\rightarrow \text{B)}} = \arg\left(\widetilde{\chi}^{\text{(A}\rightarrow \text{B)}}\right) \). We can use the relations shown in Eq. (\ref{eq:orthnormality}) to rearrange the frequency response
\begin{equation}
\widetilde{\chi}^{\text{(A}\rightarrow \text{B)}} =\Bigg[ \left( V_+^\text{(B)} {V_+^\text{(A)}}^* + \varepsilon \left|V_+^\text{(B)}\right|^2\right) \chi_+ - \left(V_+^\text{(B)} {V_+^\text{(A)}}^* - \varepsilon \left( 1- \left|V_+^\text{(B)}\right|^2\right)\right) \chi_- \Bigg].
\end{equation}

\begin{figure}[hb!]
    \centering
    \includegraphics[scale=0.77]{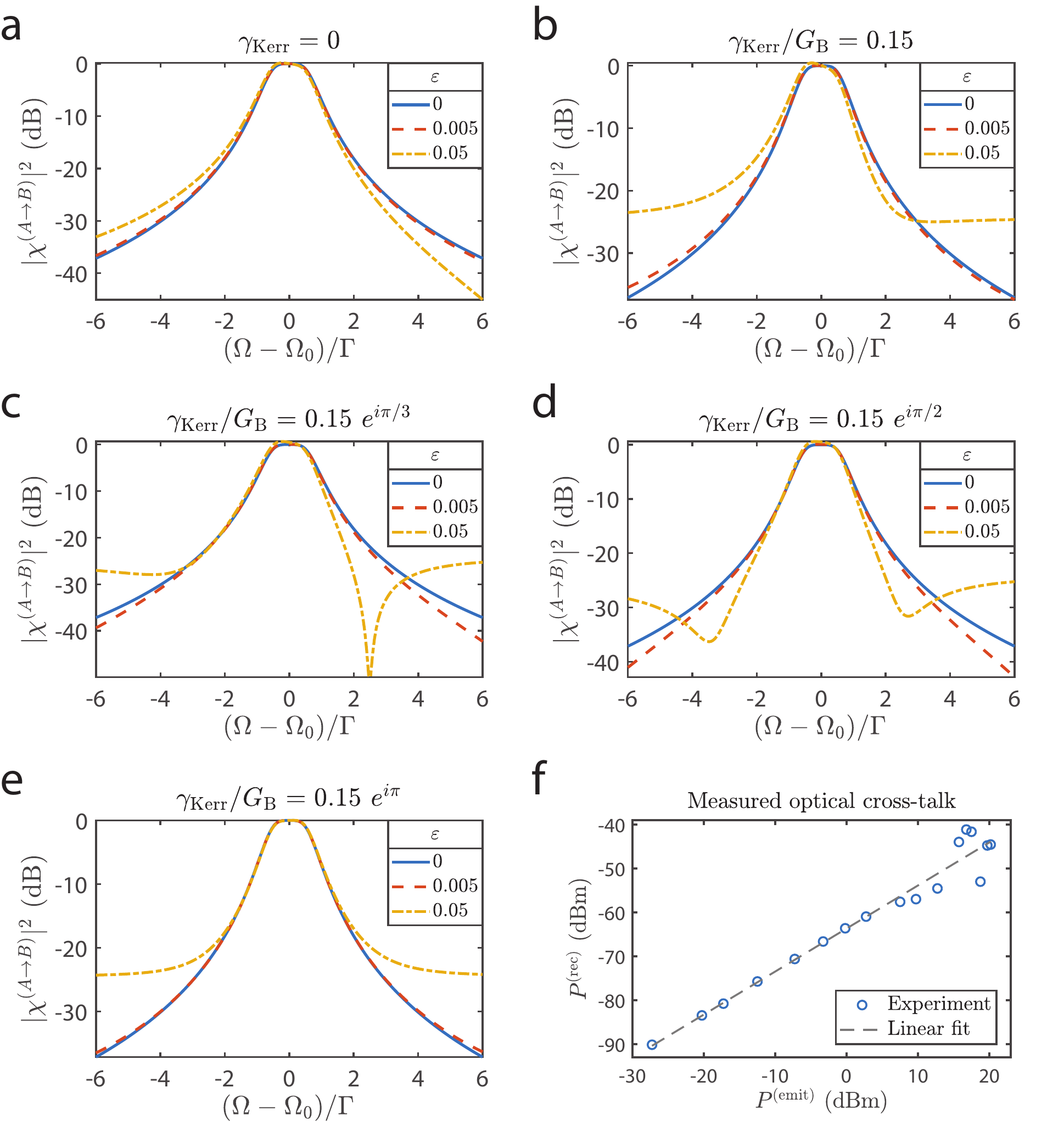}
    \caption{
    \textbf{(a)} Calculated two-pole filter shapes of a PPER operation, including different amounts of optical cross-talk $\varepsilon$.
    \textbf{(b-e)} Calculated two-pole filter shapes of a PPER operation including optical cross-talk $\varepsilon$ and Kerr-induced four-wave-mixing, showing how different relative phases between the Brillouin and Kerr nonlinearities result in different line shapes.
    The calculations assume $\Omega_0/\Gamma = 1000$ and $\mu=\Gamma/2$.
    \textbf{(f)} Measured cross-talk in a two-pole PPER device, showing $-60$ dB of optical power leaking from the `emit' to the `receive' waveguides.
    }
    \label{fig:xtalk}
\end{figure}

We see that in the case of no cross-talk $\left(\varepsilon=0\right)$ we recover the result from Eq. (\ref{eq:freq_res_asymm_2}). Assuming two identical acoustic modes, such that $V_+^\text{(B)} {V_+^\text{(A)}}^*=|V_+^\text{(B)}|^2=1/2$, we have
\begin{equation}
\widetilde{\chi}^{\text{(A}\rightarrow \text{B)}} =\frac{1}{2}\Big[ \left( 1 + \varepsilon \right) \chi_+ - \left(1 - \varepsilon \right) \chi_- \Big],
\label{eq:xtalk}
\end{equation}
revealing the asymmetrical response as a result of the cross-talk, even in the case of a symmetric device. Fig. \ref{fig:xtalk}(a) shows how the filter line shape acquires some asymmetry as the optical cross-talk becomes non-negligible. This line shape can be slightly altered when considering a phase shift induced in the coupling, equivalent to a complex-valued $\varepsilon$.

\vspace{2mm}

We can further take into account optical nonlinearity resulting from Kerr-induced four-wave-mixing in waveguide B. 
Analyzing one of the optical sidebands in waveguide B, we can write the contribution of both Brillouin and Kerr nonlinearities \cite{shin2013tailorable}
\begin{equation}
    \frac{\partial a_{-1}^\text{(B)}}{\partial z} \propto -i     \left[\frac{G_\text{B} \Gamma}{4} \widetilde{\chi}^{\text{(A}\rightarrow \text{B)}} + 2 \varepsilon \gamma_\text{Kerr} \right] \left|{a}^\text{(A)}_{0}(0)\right|^2 a_0^\text{(B)},
\end{equation}
where $G_\text{B}=4|g|^2/(\hbar \omega \Gamma v^2)$ is the Brillouin  gain coefficient \cite{kharel2016_Hamiltonian}, $\Gamma$ the acoustic dissipation rate, and $\widetilde{\chi}^{\text{(A}\rightarrow \text{B)}}$ is the Brillouin frequency response from Eq. (\ref{eq:xtalk}). $\gamma_\text{Kerr}$ is the Kerr coefficient, and $\varepsilon$ the optical cross-talk between the waveguides. We are assuming the Kerr nonlinearity has no frequency dependence over the bandwidth of interest, as it is typically much wider band ($\sim$500 GHz) compared to the Brillouin frequency response ($\sim$3 MHz). In the silicon structures we are studying in this work, the Kerr coefficient is an order of magnitude smaller than that of the Brillouion gain \cite{kittlaus2016large}.
Figs. \ref{fig:xtalk}(b-e) show calculated filter line shapes when including optical cross-talk and four-wave-mixing, assuming $|\gamma_\text{Kerr}/G_\text{B}| = 0.15$. Since these are coherent processes, varying the phase between the two nonlinear contributions will yield different line shapes. This phase difference can be a result of the optical coupling mechanism leading to the cross-talk, as well as the phase difference between the optical fields in waveguides A and B. 
The Kerr coefficient has a small magnitude compared to the Brillouin contribution, hence the four-wave-mixing results in a small change to the line shape at the center of the filter pass-band, and appreciable deviation is seen only at frequencies outside the half-maximum frequency of the filter.

We quantify the optical cross-talk in a two-pole PPER device by injecting an optical tone with power $P^\text{(emit)}$ into waveguide A, and measuring the optical power $P^\text{(rec)}$ coming out of waveguide B.
Fig. \ref{fig:xtalk}(f) presents the measured optical cross-talk, showing $-60$ dB of optical power leakage. The small value suggests that in the PPER two-pole device demonstrated in this work the effects of cross-talk will be negligible.

\twocolumngrid
\bibliographystyle{unsrtnat}
\bibliography{IMPPER_exp}

\end{document}